\documentclass[a4paper]{aa}  
\usepackage{graphicx}
\usepackage{txfonts}
\usepackage{natbib}
\usepackage{longtable}
\bibpunct{(}{)}{;}{a}{}{,}

\newcommand{\kms}{\,km\,s$^{-1}$}

\begin{document}
   \title{Interstellar reddening towards six small areas in Puppis-Vela\thanks{Table\,\ref{photometry}
   is only available in electronic form at {\tt http://www.aanda.org} and at the CDS via anonymous ftp 
   to {\tt cdsarc.u-strasbg.fr} (130.79.128.5) or via {\tt http://cdsweb.u-strasbg.fr/cgi-bin/qcat?J/A+A/}}
   \fnmsep\thanks{Based on observations collected at the European Southern Observatory (ESO, 
   La Silla, Chile)}}
   
   \titlerunning{Six small areas in Puppis-Vela}

   \author{G. A. P. Franco}

   \institute{Departamento de F\'{\i}sica - ICEx - UFMG, Caixa Postal 702,
30.123-970 - Belo Horizonte - MG, Brazil\\
              \email{franco@fisica.ufmg.br}
               }

   \date{Received February 09, 2012; accepted May 01, 2012}

  \abstract
   {The line-of-sight towards Puppis-Vela contains some of the most interesting and elusive
   objects in the solar neighbourhood, including the Gum nebula, the $IRAS$ Vela shell, the 
   Vela SNR, and dozens of cometary globules.}
   {We investigate the distribution of the interstellar dust towards six small volumes of the sky in the
   region of the Gum nebula.}
   {New high-quality four-colour $uvby$ and H$\beta$ Str\"omgren photometry obtained for 352 stars 
   in six selected areas of Kapteyn and complemented with data obtained in a previous investigation 
   for two of these areas, were used to estimate the colour excess and distance to these objects. The 
   obtained colour excess $versus$ distance diagrams, complemented with other information, when 
   available, were analysed in order to infer the properties of the interstellar medium permeating 
   the observed volumes.}
   {On the basis of the overall standard deviation in the photometric measurements, we estimate that
   colour excesses and distances are determined with an accuracy of $0\fm010$ and better than
   30\%, respectively, for a sample of 520 stars. A comparison with 37 stars in common with the new
   $Hipparcos$ catalogue attests to the high quality of the photometric distance determination. The
   obtained colour excess $versus$ distance diagrams testify to the low density volume towards the
   observed lines-of-sight. Very few stars out to distances of 1\,kpc from the Sun have colour 
   excesses larger than E$(b-y) = 0\fm1$.}
   {In spite of the low density character of the interstellar medium towards the Puppis-Vela direction, 
   the obtained reddening as a function of the distance indicates that two or more interstellar structures 
   are crossed towards the observed lines-of-sight. One of these structures may be associated with the 
   very low density ``wall'' of the Local Cavity, which has a distance of 100 -- 150\,pc from the Sun. 
   Another structure might be related to the Gum nebula, and if so, its front face would be located 
   at $\sim$350\,pc from the Sun.}


   \maketitle
%
%________________________________________________________________

\section{Introduction}

Early investigations proposed that the Sun lies in an irregularly shaped largely evacuated volume 
of gas with a minimum radius of $\sim$50\,pc and a maximum radius of 150--200\,pc (known as ``the 
Local Cavity''). A review of the properties of this low density volume may be found in \citet{FRS11}. 
Nevertheless, high resolution \ion{Na}{i}\,D absorption line studies \citep[see for instance,][]{WLV10} 
and photometric analyses \citep[e.g.,][]{RCA11} have revealed a {\it tunnel} of low 
interstellar gas density towards the Galactic longitude $l \sim 260\degr$, extending to at least 
250\,pc from the Sun. This line-of-sight coincides with the direction of Puppis-Vela ($l=245\degr$ to 
275{\degr}, $b = -15\degr$ to $+10\degr$), which is known to host some of the most interesting 
astronomical objects in the solar neighbourhood. A detailed description of the major features 
observed out to a distance of $\sim$2\,kpc is provided by \citet{pettersson08}. However, since our 
main interest in this investigation is the nearest 1\,kpc, the objects within this volume are 
briefly introduced. 

One of these features is the Gum nebula discovered by \citet{GUM52} during an H$\alpha$ survey of 
the southern Milky Way. It is the largest feature in this region and probably the most controversial 
of them. The Gum nebula appears as a spherical shell of ionised gas with an apparent diameter 
of $\approx$36\degr, centred on ($l,b$) = (258\degr,$-2$\degr) \citep{CS83}.
Slightly to the southeast, but still in the direction of the Gum nebula, \citet{Sahu92} identified an
extended ring-like dust structure in the {\it IRAS} emission maps. This structure is known as
the {\it IRAS} Vela shell (IVS), centred on ($l,b$) = (263\degr,$-7$\degr), and about 
15\degr\ in diameter.

The Vela supernova remnant, supposedly one of the closest SNR to us, is another interesting feature. 
It is found in the same direction as the Gum nebula, centred on ($l,b$) = ($263\fdg9$,$-3\fdg3$) and 
is about 8\degr\ in diameter.

%--------------------------- Figure 1 ----------------------------------------
\begin{figure*}
\centering
\includegraphics[width=\hsize]{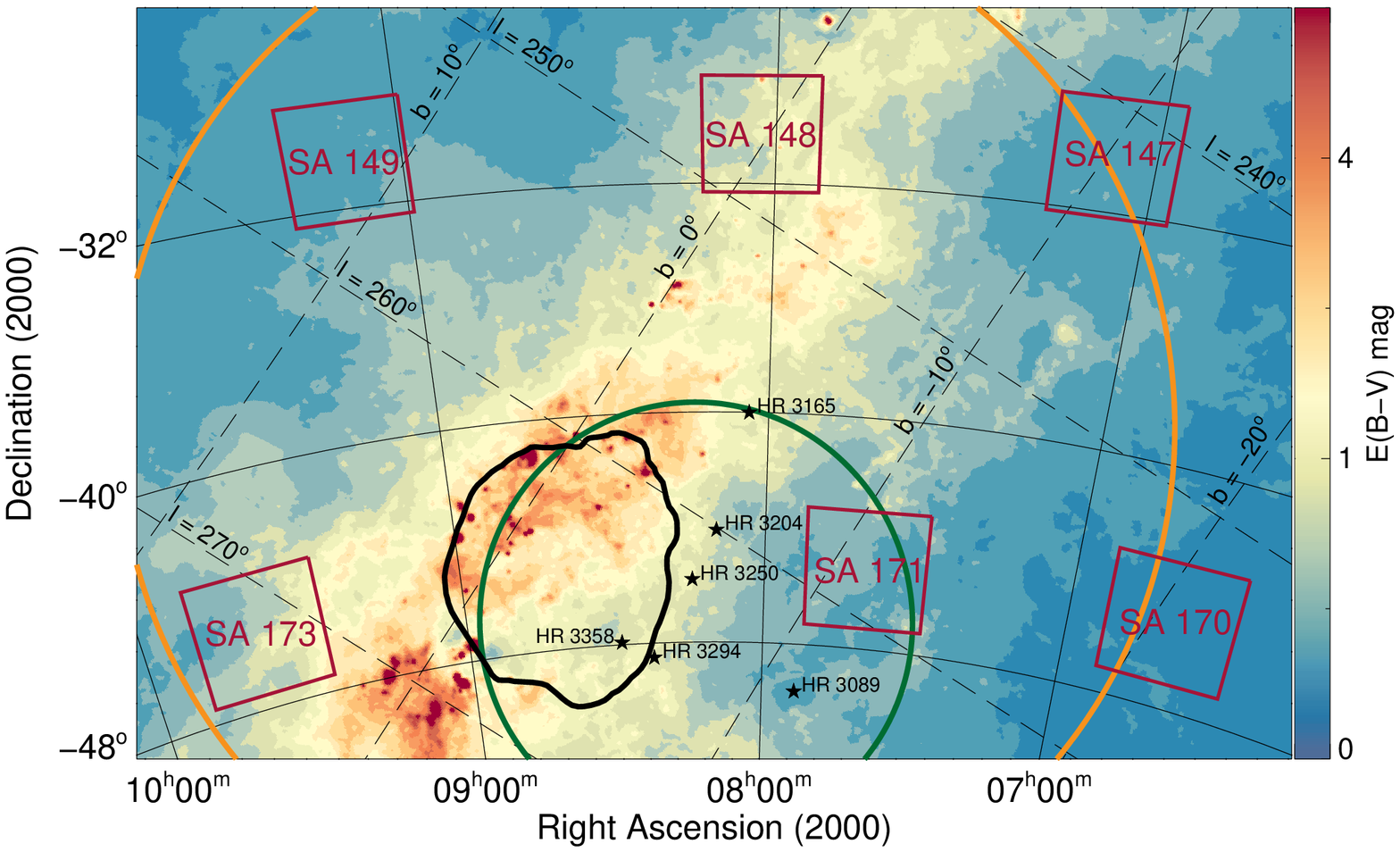}
\caption{Detail of the investigated region. The reddening image was retrieved from the all-sky
extinction map published by \citet{SFD98}. The borders of the areas observed photometrically 
are indicated by the red boxes. The approximate Gum nebula border is indicated by the large
(orange) circle of radius $\sim$\,18\degr\ centred at $\alpha_{2000} = 8^{\mathrm h}
21^{\mathrm m} 12\fs74$, $\delta_{2000} = -40\degr 12^{\prime}36\farcs7$ ($l=258\degr$, 
$b=-2\degr$), while the small (green) circle indicates the {\it IRAS}-Vela shell border, with a radius of 
$\sim\,7\fdg5$, centred at $\alpha_{2000} = 8^{\mathrm h}13^{\mathrm m} 33\fs52$, $\delta_{2000} 
= -47\degr07^{\prime} 58\farcs2$ ($l=263\degr$, $b=-7\degr$) \citep[adopted from][]{Sahu92}, and 
the black contour indicates the approximate border of the Vela SNR \citep[adopted from][]{CS00}. 
The most intense absorption regions ($ 255\degr \le l \le 275\degr$, $b = \pm 5\degr$) are likely 
associated with the Vela Molecular Ridge. The ``stars'' indicate the lines-of-sight to six stars observed 
spectroscopically, which are respectively identified by their HR numbers \citep{HD82}. (A colour 
version of this figure is available in the online journal).}
\label{sfd_map}
\end{figure*}
%-------------------------------------------------------------------------------

In addition, a few dozens of cometary globules (CG) have been discovered in the region 
of the Gum nebula distributed in a nearly circular pattern \citep{HB76, sandqvist76, ZNR83, reipurth83}. 
These all have compact dense heads that are opaque to the background starlight. On the side 
pointing towards the centre of the nebula, they are well-defined with very sharp edges often 
exhibiting narrow bright rims, while on the opposite side they have protruding slightly luminous tails.

The distances and origins of these interstellar features are controversial and debated issues. For
instance, \citet{WGO01} suggested that the diffuse gas composing the Gum nebula and the
cometary globules system form a single expanding shell with a maximum radius of 14\degr. These
authors also claimed that no evidence was found that the IVS is a separate expanding shell, and
suggested that it is a density enhancement in the neutral Gum nebula shell. As the result of their 
kinematical model, they suggested that the Gum nebula can be more accurately described as a 
non-uniformly expanding shell in which the front face expands faster than the back face, centred 
at 500\,pc from the Sun. In this model, the front face would be at about 350\,pc from us. Other 
authors have suggested values for the distance to the centre of the Gum nebula that varies from 290\,pc
\citep{FR90, KN00} to 800\,pc \citep{SB93}. Those latter authors proposed that the IVS encompasses
the Vela OB2 association at a distance of about 450\,pc \cite[more recent estimates locates
this association slightly closer to the Sun, see e.g.,][]{ZHB99,JNW09}. In the \citeauthor{SB93}'s
scenario, the Gum nebula, and IVS are two independent, but not interacting, entities. The front
faces should be at $\sim$390\,pc and $\sim$540\,pc from the Sun, respectively for the IVS and the
Gum nebula.

The distance estimate to the centre of the Vela SNR also suffers from a large uncertainty, ranging 
from 250$\pm$30\,pc \citep{CSD99} to the assumed 500\,pc by \citet{milne68}, or even more. 
Nevertheless, the best estimate seems to be the one obtained by VLBI parallax measure 
\citep{DLR03}, which provides a distance of 287$\pm$19\,pc.

The next structure is less controversial. Behind the Gum nebula, there is a large molecular cloud 
complex known as the Vela Molecular Ridge (VMR) \citep{MMT88, MM91}. This giant molecular 
complex extends roughly 20\degr\ in the sky and comprises four main clouds designated $A$ to $D$ 
by \citet{MM91}. These authors suggested distances of about 1\,kpc to components $A$, $C$, 
and $D$, and about 2\,kpc to component $B$. Each of these main clouds have masses exceeding 
10$^5$\,M\sun. \citet{LLN92} investigated the distance to this complex, finding that clouds 
$A$, $C$, and $D$ are likely at $700 \pm 200$\,pc, whereas cloud $B$ appears to be at 
$\sim$2\,kpc. These authors also identified dozens of objects that could be classified as Class I 
sources, reinforcing the suggestion made by \citet{MM91} that this large complex is an active 
region of star formation. Furthermore, \citet{NAB09} found that 2\% of the mass of the molecular gas 
comprising this complex is in cold cores in a range of evolutionary stages. In addition, two dozen
embedded stellar clusters are believed to be associated to this giant molecular complex 
\citep[e.g.,][]{BDB03, DBS03}.

This paper introduces the analysis of the interstellar reddening obtained from high-quality Str\"omgren
photometry of stars belonging to six selected areas of Kapteyn with lines-of-sight towards the volume
encompassing the Gum nebula and its neighbourhood structures. 

\section{Observational data}

The observations were performed with the Str\"omgren Automatic Telescope
(SAT) of the Copenhagen Astronomical Observatory at La Silla, Chile. 
The telescope was equipped with a six-channel {\it uvby--}$\beta$ 
spectrograph-photometer \citep{FN}, which allows simultaneous measurements 
for the {\it uvby} passbands, or simultaneous measurements for the narrow 
and wide passbands used to define the $\beta$ index. The procedure used to 
collect the measurements as well as the method employed to reduce the data, 
such as extinction correction and transformations to the {\it uvby} and to 
the $\beta$ standard systems, were previously described by \citet{FR94}. 

The observing list is based on the Potsdam Spektral-Durchmusterung 
\citep{BB}. All stars brighter than $m_{pg} =11\fm0$, earlier than G0 and
belonging to the selected areas SA\,147, SA\,148, SA\,149, and SA\,170
were chosen for observation. In addition, the previous observed
sample for selected areas SA\,171 and SA\,173 \citep{FR88b} were
complemented to fulfil the same spectral coverage as the former four areas.
All the observed areas have sizes of 16 square degrees in the Potsdam 
Spektral-Durchmusterung.

A total of 352 stars were observed at least twice in both modes, $uvby$ and 
H$\beta$. The $uvby$ measurements were used to obtain for each star the colour 
index ($b-y$), the colour index differences $m_1 = [(v-b) - (b-y)]$ and $c_1 =
[(u-v) - (v-b)]$ on the standard $uvby$ system \citep{CB70, GOS76}, and the visual 
photometric $V$ magnitude on the Johnson system. The H$\beta$ index is on the
standard $\beta$ system \citep{CM66}. The obtained results are introduced in 
Table\,\ref{photometry} (available electronically only). The estimated overall rms errors 
for one observation of one star are $0\fm006$, $0\fm004$, $0\fm005$, $0\fm008$, 
and $0\fm010$ in $V$, $b-y$, $m_1$, $c_1$, and H$\beta$, respectively. 

There are 191 stars in the previously observed sample for SA\,171 and SA\,173 \citep{FR88b},
which combined with the one introduced here makes a total of 543 stars for the entire set of 
selected areas.

Figure\,\ref{sfd_map} depicts a portion of the large region containing the Gum nebula. The borders 
of the six observed selected areas of Kapteyn are superposed on a reddening E$(B-V)$ 
image retrieved from the all-sky reddening map published by \citet[][hereafter SFD98]{SFD98}.
For reference, the approximate borders of the Gum nebula, the IVS, and the Vela SNR are 
indicated, respectively, by the larger and smaller circles and the irregular contour. Two of 
these areas, SA\,147 and SA\,170, are only partially inside the Gum nebula's border, while SA\,171 
also probes the line-of-sight towards the IVS. The remaining three areas probe different parts of 
the Gum nebula's volume. The ``star'' signs indicate the lines-of-sight towards six stars for which 
interstellar \ion{Na}{i}\,D lines spectra are also analysed (see Sect.\,\ref{sodio}).

\onllongtab{1}{
\begin{longtable}{rrlrrrrrrrlr}
\caption{\label{photometry} Str\"omgren photometry of stars earlier than G0 
towards six selected areas in Puppis-Vela. The first five columns give the 
identification in the Potsdam Spektral-Durchmusterung Catalog \citep{BB}, HD 
number (when available), Michigan two-dimensional classification 
\citep{HN1978, HN1982}, and right ascension and declination for the equinox 
2000.0. Columns 6 to 9 and 11 give the computed V magnitude on the Johnson 
system, the colour index ($b-y$), $m_1$, $c_1$, and H$\beta$, respectively. 
Figures between parentheses give the standard deviation in the measurements 
in units of $0\fm001$. Columns 10 and 12 give the number of nights in which 
the star was observed in $uvby$ and H$\beta$, respectively.
} \\
\hline
\hline
\multicolumn{1}{c}{SA} & \multicolumn{1}{c}{HD} & \multicolumn{1}{c}{Spectral}
& \multicolumn{1}{c}{$\alpha_{\rm 2000}$} & \multicolumn{1}{c}{$\delta_{\rm 2000}$}
& \multicolumn{1}{c}{V} & \multicolumn{1}{c}{(b$-$y)} & \multicolumn{1}{c}{m$_{_1}$}
& \multicolumn{1}{c}{c$_{_1}$} & \multicolumn{1}{c}{n} & \multicolumn{1}{c}{H$\beta$}
 & \multicolumn{1}{c}{n} \cr
& & \multicolumn{1}{c}{type} & \multicolumn{1}{c}{(h m s)} &
\multicolumn{1}{c}{(\degr\hskip.2cm $'$\hskip.2cm $''$)} & \multicolumn{1}{c}{(mag)} &
\multicolumn{1}{c}{(mag)} & \multicolumn{1}{c}{(mag)} &
\multicolumn{1}{c}{(mag)} & & \multicolumn{1}{c}{(mag)} & \cr
\endfirsthead
\caption{Continued.} \\
\hline
\multicolumn{1}{c}{SA} & \multicolumn{1}{c}{HD} & \multicolumn{1}{c}{Spectral}
& \multicolumn{1}{c}{$\alpha_{\rm 2000}$} & \multicolumn{1}{c}{$\delta_{\rm 2000}$}
& \multicolumn{1}{c}{V} & \multicolumn{1}{c}{(b$-$y)} & \multicolumn{1}{c}{m$_{_1}$}
& \multicolumn{1}{c}{c$_{_1}$} & \multicolumn{1}{c}{n} & \multicolumn{1}{c}{H$\beta$}
 & \multicolumn{1}{c}{n} \cr
& & \multicolumn{1}{c}{type} & \multicolumn{1}{c}{(h m s)} &
\multicolumn{1}{c}{(\degr\hskip.2cm $'$\hskip.2cm $''$)} & \multicolumn{1}{c}{(mag)} &
\multicolumn{1}{c}{(mag)} & \multicolumn{1}{c}{(mag)} &
\multicolumn{1}{c}{(mag)} & & \multicolumn{1}{c}{(mag)} & \cr
\hline
\endhead
\hline
147.~~~~39  & 51858 & F(5) V         & 06 57 33& -30 21 01&  9.911 ( 7)&  0.327 ( 7)&  0.147 ( 4)&  0.439 (11)& 2 &2.625 (12)& 2 \cr
147.~~102   & 52056 & F7 V           & 06 58 27& -30 16 09&  9.150 ( 5)&  0.320 ( 2)&  0.158 ( 6)&  0.424 ( 7)& 2 &2.638 (12)& 3 \cr
147.~~104   & 52091 & A8/9 V         & 06 58 27& -30 30 09&  9.643 ( 3)&  0.195 ( 5)&  0.170 ( 9)&  0.865 ( 4)& 2 &2.789 (16)& 3 \cr
147.~~201   &       &                & 06 59 42& -31 17 20& 10.113 ( 6)&  0.175 ( 1)&  0.186 ( 0)&  0.927 ( 2)& 2 &2.828 (20)& 2 \cr
147.~~210   & 52442 & F2 IV/V        & 07 00 04& -28 10 22&  8.708 ( 1)&  0.262 ( 6)&  0.182 (12)&  0.651 (16)& 2 &2.698 (15)& 3 \cr
147.~~233   &       &                & 07 00 06& -31 34 23&  9.893 (14)&  0.224 ( 3)&  0.162 ( 4)&  0.650 ( 0)& 2 &2.734 ( 0)& 2 \cr
147.~~234   & 52517 & F7/G0 V        & 07 00 05& -31 37 23&  9.181 ( 7)&  0.341 ( 7)&  0.158 (16)&  0.408 (18)& 2 &2.625 (15)& 3 \cr
147.~~237   & 52545 & F7wG2/3        & 07 00 18& -29 24 24&  9.355 ( 4)&  0.359 ( 1)&  0.145 ( 1)&  0.345 ( 7)& 2 &2.614 ( 7)& 3 \cr
147.~~244   & 52516 & A4/5 IV        & 07 00 13& -31 08 24&  6.619 ( 5)&  0.297 ( 1)&  0.151 ( 1)&  0.416 ( 4)& 2 &2.660 ( 3)& 3 \cr
147.~~266   &       &                & 07 00 37& -29 14 27&  9.801 ( 7)&  0.383 ( 0)&  0.186 ( 0)&  0.705 (24)& 2 &2.676 ( 9)& 2 \cr
147.~~274   & 52619 & F3/5 V         & 07 00 45& -28 29 28&  6.293 ( 4)&  0.296 ( 2)&  0.159 ( 2)&  0.520 ( 6)& 2 &2.662 ( 3)& 3 \cr
147.~~277   &       &                & 07 00 40& -30 06 28& 10.215 (12)&  0.371 ( 4)&  0.166 ( 4)&  0.482 ( 9)& 2 &2.633 ( 8)& 2 \cr
147.~~299   & 52672 & F2 V           & 07 00 49& -31 06 29&  9.030 ( 0)&  0.259 ( 2)&  0.172 ( 3)&  0.739 (16)& 2 &2.699 (11)& 3 \cr
147.~~307   & 52702 & F0 V           & 07 00 57& -30 33 30&  8.559 ( 2)&  0.221 ( 5)&  0.156 ( 4)&  1.040 ( 0)& 2 &2.743 (25)& 3 \cr
147.~~323   & 52782 & A9 IV          & 07 01 09& -30 39 32&  7.405 ( 6)&  0.160 ( 1)&  0.195 ( 0)&  0.830 ( 5)& 2 &2.795 (11)& 3 \cr
147.~~329   &       &                & 07 01 17& -29 52 33& 10.509 ( 4)&  0.203 ( 9)&  0.166 (12)&  0.758 (11)& 2 &2.767 (26)& 2 \cr
147.~~355   &       &                & 07 01 24& -31 21 34&  9.671 ( 9)&  0.315 ( 3)&  0.146 ( 9)&  0.508 ( 9)& 2 &2.657 ( 3)& 2 \cr
147.~~398   & 52991 & F5 V           & 07 02 08& -28 55 40&  9.130 ( 8)&  0.320 ( 0)&  0.155 ( 4)&  0.520 ( 6)& 2 &2.655 (13)& 3 \cr
147.~~442   &       &                & 07 02 28& -30 21 43&  9.743 ( 5)&  0.281 ( 0)&  0.141 ( 4)&  0.488 (10)& 2 &2.659 (17)& 2 \cr
147.~~545   & 53376 & F3/5 V         & 07 03 29& -31 48 52&  9.325 ( 0)&  0.298 ( 2)&  0.153 ( 4)&  0.470 (10)& 2 &2.635 (19)& 2 \cr
147.~~561   & 53463 & A9/F0          & 07 03 45& -30 43 54&  9.164 ( 9)&  0.252 ( 4)&  0.155 ( 5)&  0.519 ( 2)& 2 &2.681 (21)& 2 \cr
147.~~581   &       &                & 07 04 03& -30 42 56& 10.082 ( 3)&  0.259 ( 3)&  0.156 ( 8)&  0.521 ( 0)& 2 &2.697 (12)& 2 \cr
147.~~589   &       &                & 07 04 07& -31 11 57& 10.028 ( 4)&  0.322 ( 0)&  0.150 ( 2)&  0.438 ( 2)& 2 &2.639 (10)& 2 \cr
147.~~593   &       &                & 07 04 13& -31 09 58&  9.949 ( 0)&  0.344 ( 9)&  0.160 ( 7)&  0.488 ( 9)& 2 &2.640 ( 1)& 2 \cr
147.~~597   & 53601 & F6 V           & 07 04 26& -28 53 59&  8.979 ( 6)&  0.328 ( 4)&  0.166 ( 7)&  0.415 ( 3)& 2 &2.630 ( 2)& 2 \cr
147.~~610   &       &                & 07 04 27& -30 42 00& 10.227 ( 5)&  0.334 ( 4)&  0.143 ( 6)&  0.430 ( 2)& 2 &2.632 ( 9)& 2 \cr
147.~~642   & 53698 & F3 V           & 07 04 47& -31 57 03&  7.341 ( 2)&  0.291 ( 2)&  0.146 ( 0)&  0.502 ( 3)& 2 &2.655 ( 0)& 2 \cr
147.~~648   &       &                & 07 04 54& -31 35 04&  9.989 ( 2)&  0.316 ( 2)&  0.146 ( 7)&  0.546 ( 0)& 2 &2.646 ( 4)& 2 \cr
147.~~698   &       &                & 07 05 45& -28 30 10& 10.120 ( 7)&  0.265 (11)&  0.150 (12)&  0.486 ( 4)& 2 &2.671 (23)& 2 \cr
147.~~754   &       &                & 07 06 22& -28 22 15& 10.009 ( 9)&  0.007 ( 8)&  0.107 (11)&  0.851 (16)& 2 &2.825 ( 5)& 2 \cr
147.~~776   & 54152 & F0 V           & 07 06 28& -30 16 17&  9.177 ( 8)&  0.210 ( 0)&  0.168 ( 4)&  0.727 ( 1)& 2 &2.729 ( 0)& 2 \cr
147.~~778   &       &                & 07 06 38& -28 50 18& 10.259 ( 4)&  0.286 ( 2)&  0.152 (12)&  0.524 (13)& 2 &2.676 (14)& 2 \cr
147.~~793   &       &                & 07 06 41& -30 09 18&  9.673 ( 2)&  0.256 ( 0)&  0.146 ( 4)&  0.612 ( 1)& 2 &2.696 ( 7)& 2 \cr
147.~~800   &       &                & 07 06 47& -29 54 19&  9.662 ( 6)&  0.163 ( 4)&  0.193 ( 7)&  0.940 ( 5)& 2 &2.822 ( 8)& 2 \cr
147.~~805   &       &                & 07 06 44& -30 56 19& 10.082 ( 1)&  0.324 ( 0)&  0.171 ( 2)&  0.552 (10)& 2 &2.674 ( 8)& 2 \cr
147.~~807   &       &                & 07 06 41& -31 54 19&  9.803 ( 9)&  0.331 ( 4)&  0.162 ( 4)&  0.514 ( 7)& 2 &2.663 ( 3)& 2 \cr
147.~~824   &       &                & 07 07 10& -28 24 22&  9.424 ( 4)&  0.316 ( 6)&  0.131 ( 4)&  0.394 ( 4)& 2 &2.639 (12)& 2 \cr
147.~~839   &       &                & 07 07 08& -31 09 22& 10.162 ( 4)&  0.286 ( 4)&  0.165 ( 0)&  0.504 ( 2)& 2 &2.673 ( 1)& 2 \cr
147.~~840   &       &                & 07 07 07& -31 16 22& 10.348 ( 1)&  0.367 ( 5)&  0.160 ( 2)&  0.413 ( 1)& 2 &2.620 ( 6)& 2 \cr
147.~~869   &       &                & 07 07 31& -31 23 26& 10.201 ( 4)&  0.204 ( 4)&  0.174 ( 4)&  0.698 ( 7)& 2 &2.768 (11)& 2 \cr
147.~~874   & 54472 & G8/K0 III + A3 & 07 07 44& -28 59 27&  8.405 ( 8)&  0.421 ( 4)&  0.184 ( 3)&  0.663 (12)& 2 &2.646 ( 8)& 2 \cr
147.~~883   &       &                & 07 07 49& -29 31 28& 10.412 ( 2)&  0.285 ( 0)&  0.152 ( 4)&  0.570 (21)& 2 &2.692 ( 2)& 2 \cr
147.~~889   & 54499 & A3 III/IV      & 07 07 42& -31 46 27&  9.881 ( 0)&  0.110 ( 4)&  0.186 ( 7)&  1.091 ( 0)& 2 &2.852 (59)& 2 \cr
147.~~892   &       &                & 07 07 55& -29 17 29& 10.195 ( 7)&  0.317 ( 0)&  0.150 ( 6)&  0.418 (11)& 2 &2.647 ( 5)& 2 \cr
147.~~897   &       &                & 07 07 50& -30 53 28& 10.000 (11)&  0.284 ( 4)&  0.129 ( 7)&  0.425 ( 9)& 2 &2.639 (20)& 2 \cr
147.~~920   &       &                & 07 08 01& -31 29 30&  8.986 ( 4)&  0.220 ( 0)&  0.160 ( 0)&  0.734 ( 0)& 2 &2.726 (12)& 2 \cr
147.~~927   &       &                & 07 08 05& -31 57 31&  9.636 ( 4)&  0.176 ( 6)&  0.171 ( 2)&  0.736 (17)& 2 &2.758 ( 0)& 2 \cr
147.~~930   &       &                & 07 08 17& -29 57 32& 10.149 ( 1)&  0.216 ( 0)&  0.155 ( 2)&  0.651 (33)& 2 &2.753 (33)& 2 \cr
147.~~931   & 54626 & F5 IV/V        & 07 08 13& -31 25 31&  7.059 ( 2)&  0.299 ( 2)&  0.158 ( 1)&  0.516 ( 6)& 2 &2.658 ( 1)& 2 \cr
147.~~943   &       &                & 07 08 30& -29 49 33&  9.732 ( 6)&  0.314 ( 3)&  0.127 ( 4)&  0.379 ( 3)& 2 &2.627 ( 4)& 2 \cr
147.~~946   &       &                & 07 08 40& -28 25 35& 10.332 (12)&  0.178 ( 6)&  0.159 ( 4)&  0.830 (12)& 2 &2.781 ( 5)& 2 \cr
147.~~948   &       &                & 07 08 37& -29 30 34&  9.942 ( 4)&  0.303 ( 1)&  0.143 (12)&  0.518 (26)& 2 &2.661 ( 0)& 2 \cr
147.~~991   &       &                & 07 09 03& -30 53 38&  9.683 ( 6)&  0.226 ( 0)&  0.236 ( 4)&  0.733 ( 7)& 2 &2.749 ( 4)& 2 \cr
147.1045    &       &                & 07 09 49& -29 35 44& 10.494 ( 2)&  0.178 ( 1)&  0.181 ( 2)&  0.870 ( 2)& 2 &2.772 (14)& 2 \cr
147.1048    &       &                & 07 09 47& -30 12 44& 10.493 ( 4)&  0.126 (14)&  0.201 (19)&  0.911 ( 8)& 2 &2.853 ( 8)& 2 \cr
147.1049    &       &                & 07 09 45& -30 49 44& 10.076 ( 7)&  0.276 ( 4)&  0.171 ( 6)&  0.546 ( 2)& 2 &2.688 (14)& 2 \cr
147.1054    &       &                & 07 09 57& -28 36 45&  9.813 ( 1)&  0.339 ( 9)&  0.149 ( 7)&  0.446 ( 1)& 2 &2.642 (14)& 2 \cr
147.1078    &       &                & 07 10 03& -30 41 47&  9.227 ( 9)&  0.354 ( 0)&  0.152 ( 0)&  0.417 ( 9)& 2 &2.614 ( 2)& 2 \cr
147.1090    &       &                & 07 10 14& -31 12 48&  9.889 ( 2)&  0.378 ( 9)&  0.186 (12)&  0.354 ( 4)& 2 &2.607 (14)& 2 \cr
147.1099    &       &                & 07 10 23& -29 58 49&  9.744 ( 0)&  0.345 ( 2)&  0.143 ( 0)&  0.499 (11)& 2 &2.625 ( 8)& 2 \cr
147.1100    & 55122 & A7 V           & 07 10 23& -30 03 49&  7.755 ( 9)&  0.121 ( 2)&  0.192 ( 0)&  0.895 ( 2)& 2 &2.827 ( 0)& 2 \cr
147.1113    & 55143 & F6 V           & 07 10 31& -29 20 50&  7.912 ( 6)&  0.301 ( 0)&  0.157 ( 5)&  0.494 (10)& 2 &2.662 ( 4)& 2 \cr
147.1129    &       &                & 07 10 35& -30 12 51&  9.690 ( 2)&  0.314 ( 2)&  0.156 ( 2)&  0.460 ( 9)& 2 &2.657 ( 8)& 2 \cr
147.1148    & 55215 & A7 V           & 07 10 44& -31 10 52&  9.298 ( 2)&  0.125 ( 0)&  0.170 ( 0)&  1.014 ( 2)& 2 &2.833 ( 2)& 2 \cr
147.1176    &       &                & 07 11 14& -29 13 56&  9.681 ( 4)&  0.230 ( 0)&  0.173 ( 1)&  0.624 ( 1)& 2 &2.736 ( 4)& 2 \cr
147.1185    &       &                & 07 11 18& -29 50 57&  9.687 ( 6)&  0.312 ( 4)&  0.166 ( 9)&  0.411 (12)& 2 &2.626 (33)& 2 \cr
147.1215    &       &                & 07 11 30& -31 56 59&  9.386 ( 7)&  0.284 ( 2)&  0.143 ( 4)&  0.430 (27)& 2 &2.658 ( 1)& 2 \cr
147.1218    & 55447 & F7/8 V         & 07 11 42& -29 49 00&  7.808 ( 1)&  0.430 ( 3)&  0.222 ( 0)&  0.623 ( 2)& 2 &2.638 ( 5)& 2 \cr
147.1244    &       &                & 07 12 01& -29 33 03& 10.410 ( 7)&  0.221 (12)&  0.205 ( 9)&  0.728 ( 0)& 2 &2.755 ( 2)& 2 \cr
147.1254    & 55568 & A8 III/IV      & 07 12 09& -30 49 04&  6.094 ( 0)&  0.167 ( 1)&  0.178 ( 0)&  0.726 ( 0)& 2 &2.768 ( 1)& 2 \cr
147.1255    &       &                & 07 12 03& -30 52 03& 10.283 ( 7)&  0.031 ( 4)&  0.154 ( 3)&  1.088 (19)& 2 &2.882 (15)& 2 \cr
147.1258    &       &                & 07 12 00& -31 54 03& 10.494 (16)&  0.161 ( 2)&  0.185 ( 8)&  0.937 ( 0)& 2 &2.817 ( 4)& 2 \cr
147.1289    & 55694 & F5/6 V         & 07 12 32& -31 05 08&  8.907 (19)&  0.359 ( 0)&  0.157 ( 4)&  0.362 (12)& 2 &2.620 ( 2)& 2 \cr
147.1290    &       &                & 07 12 32& -31 17 07& 10.373 ( 6)&  0.286 ( 7)&  0.160 ( 2)&  0.462 ( 0)& 2 &2.672 (14)& 2 \cr
147.1324    &       &                & 07 13 08& -29 13 12&  9.688 (10)&  0.124 ( 6)&  0.194 (12)&  0.879 ( 0)& 2 &2.827 ( 4)& 2 \cr
147.1380    &       &                & 07 13 36& -31 46 17& 10.310 ( 9)&  0.214 ( 0)&  0.159 ( 0)&  0.685 ( 3)& 2 &2.732 ( 3)& 2 \cr
147.1389    &       &                & 07 14 01& -29 36 20& 10.278 ( 1)&  0.186 ( 5)&  0.164 ( 0)&  0.889 (13)& 2 &2.755 ( 7)& 2 \cr
147.1407    &       &                & 07 14 05& -32 05 21& 10.303 ( 9)&  0.234 ( 2)&  0.159 (10)&  0.575 ( 0)& 2 &2.720 ( 6)& 2 \cr
147.1412    &       &                & 07 14 13& -31 33 22& 10.148 ( 9)&  0.241 ( 2)&  0.160 ( 0)&  0.639 ( 4)& 2 &2.703 ( 7)& 2 \cr
147.1421    &       &                & 07 14 33& -29 00 24& 10.021 (16)&  0.190 ( 2)&  0.159 ( 1)&  0.921 ( 6)& 2 &2.724 (68)& 3 \cr
147.1459    &       &                & 07 14 50& -31 24 27& 10.093 ( 4)&  0.307 (10)&  0.147 ( 9)&  0.673 ( 9)& 2 &2.675 ( 4)& 2 \cr
147.1469    & 56258 & F5 V           & 07 15 05& -30 11 29&  8.715 (21)&  0.333 ( 5)&  0.163 ( 2)&  0.375 ( 5)& 2 &2.628 (11)& 2 \cr
147.1476    &       &                & 07 15 22& -28 41 31&  9.370 (13)&  0.321 ( 3)&  0.170 ( 1)&  0.534 ( 5)& 2 &2.665 (12)& 2 \cr
147.1509    &       &                & 07 15 37& -31 32 33&  9.880 (18)&  0.200 ( 3)&  0.173 ( 8)&  0.837 (23)& 2 &2.743 ( 7)& 2 \cr
148.~~~~16  &       &                & 07 53 19& -29 18 35& 10.011 ( 2)&  0.235 ( 1)&  0.148 ( 1)&  0.683 (19)& 2 &2.721 ( 8)& 3 \cr
148.~~~~24  & 64595 & F2 V           & 07 53 24& -29 49 36&  8.075 ( 4)&  0.222 ( 3)&  0.165 ( 3)&  0.791 ( 5)& 3 &2.716 ( 1)& 3 \cr
148.~~~~49  &       &                & 07 53 50& -28 58 39& 10.101 ( 2)&  0.310 ( 1)&  0.134 ( 3)&  0.461 ( 4)& 2 &2.642 ( 8)& 3 \cr
148.~~~~70  &       &                & 07 53 49& -31 27 40& 10.357 ( 2)&  0.225 ( 0)&  0.189 ( 0)&  0.625 ( 3)& 2 &2.729 ( 6)& 3 \cr
148.~~~~91  & 64757 & A9 V           & 07 54 12& -29 32 42&  9.001 ( 3)&  0.205 ( 2)&  0.167 ( 2)&  0.745 (12)& 3 &2.739 (17)& 3 \cr
148.~~~~98  &       &                & 07 54 08& -31 05 42& 10.485 ( 9)&  0.247 ( 1)&  0.155 ( 6)&  0.534 (22)& 2 &2.686 ( 4)& 3 \cr
148.~~100   & 64738 & F0 IV          & 07 54 06& -31 43 42&  8.250 ( 6)&  0.175 ( 1)&  0.189 ( 5)&  0.812 ( 6)& 3 &2.786 (15)& 3 \cr
148.~~116   & 64801 & F3 V           & 07 54 22& -30 33 44&  9.716 ( 2)&  0.285 ( 3)&  0.159 ( 5)&  0.476 ( 9)& 2 &2.670 (14)& 3 \cr
148.~~120   &       &                & 07 54 33& -28 37 45&  9.738 ( 6)&  0.324 ( 6)&  0.164 ( 5)&  0.571 ( 9)& 2 &2.664 (10)& 3 \cr
148.~~139   & 64862 & A9 V           & 07 54 46& -30 19 47& 10.067 ( 0)&  0.197 ( 3)&  0.174 ( 0)&  0.757 ( 7)& 2 &2.769 (24)& 3 \cr
148.~~143   &       &                & 07 54 57& -28 44 48& 10.078 ( 3)&  0.151 ( 6)&  0.174 ( 7)&  0.988 (12)& 2 &2.800 (12)& 2 \cr
148.~~177   & 64973 & F3 V           & 07 55 20& -28 53 51&  8.997 ( 1)&  0.317 ( 4)&  0.164 ( 5)&  0.399 (14)& 3 &2.638 (16)& 2 \cr
148.~~243   & 65087 & B2/3 II/III    & 07 55 57& -28 31 56&  9.516 ( 0)&  0.143 ( 3)& -0.025 (10)&  0.279 (15)& 4 &2.617 ( 8)& 2 \cr
148.~~252   &       &                & 07 55 48& -31 52 55& 10.498 ( 0)&  0.221 ( 2)&  0.071 ( 1)&  1.230 ( 0)& 2 &2.815 (22)& 2 \cr
148.~~262   & 65107 & A9 V           & 07 55 58& -30 34 56&  9.922 ( 2)&  0.210 ( 7)&  0.173 ( 9)&  0.690 ( 0)& 2 &2.740 ( 0)& 2 \cr
148.~~263   &       &                & 07 55 57& -30 48 56& 10.540 ( 7)&  0.162 (16)&  0.000 (11)&  0.508 ( 2)& 2 &2.688 ( 9)& 2 \cr
148.~~290   &       &                & 07 56 25& -29 25 59&  9.744 ( 7)&  0.241 ( 2)& -0.066 (15)& -0.033 ( 4)& 3 &2.514 ( 4)& 2 \cr
148.~~293   &       &                & 07 56 22& -30 27 59& 10.018 ( 2)&  0.270 ( 4)&  0.149 ( 8)&  0.514 (17)& 2 &2.679 (42)& 2 \cr
148.~~304   & 65232 & F6 V           & 07 56 28& -30 28 00&  9.300 ( 3)&  0.335 ( 1)&  0.147 ( 5)&  0.407 ( 6)& 3 &2.611 ( 2)& 2 \cr
148.~~318   &       &                & 07 56 45& -28 45 02& 10.126 ( 9)&  0.218 ( 9)&  0.163 (14)&  0.786 ( 5)& 2 &2.740 ( 2)& 2 \cr
148.~~333   &       &                & 07 56 56& -29 09 03& 10.335 ( 2)&  0.257 ( 7)&  0.160 (16)&  0.643 ( 9)& 2 &2.705 ( 7)& 2 \cr
148.~~338   &       &                & 07 56 54& -29 54 03&  9.947 ( 7)&  0.242 ( 1)&  0.135 ( 0)&  0.657 ( 3)& 2 &2.692 ( 7)& 2 \cr
148.~~352   & 65311 & F3 IV/V        & 07 57 02& -29 15 04&  9.128 ( 0)&  0.589 ( 1)&  0.115 ( 3)&  1.225 (15)& 2 &2.684 ( 2)& 2 \cr
148.~~353   & 65312 & +G             & 07 57 00& -29 43 04&  9.552 ( 9)&  0.281 ( 0)&  0.146 ( 6)&  0.455 ( 7)& 3 &2.660 ( 8)& 2 \cr
148.~~388   & 65399 & F3 V           & 07 57 28& -28 20 07&  9.506 ( 0)&  0.298 ( 2)&  0.146 ( 2)&  0.542 ( 3)& 2 &2.676 ( 2)& 2 \cr
148.~~413   &       &                & 07 57 37& -29 36 08& 10.103 ( 4)&  0.348 ( 0)&  0.149 ( 2)&  0.411 ( 2)& 2 &2.619 ( 2)& 2 \cr
148.~~451   & 65512 & F3 V           & 07 57 48& -31 54 10&  8.511 ( 1)&  0.299 ( 5)&  0.162 ( 7)&  0.636 (12)& 3 &2.686 ( 0)& 2 \cr
148.~~454   &       &                & 07 58 03& -28 56 12&  9.813 ( 1)&  0.332 ( 4)&  0.145 ( 7)&  0.408 (21)& 2 &2.633 (10)& 2 \cr
148.~~467   &       &                & 07 58 08& -29 10 12& 10.221 ( 7)&  0.225 ( 0)&  0.176 ( 7)&  0.804 (12)& 2 &2.747 ( 4)& 2 \cr
148.~~489   &       &                & 07 58 06& -32 06 13&  9.144 ( 4)&  0.276 ( 4)& -0.079 (19)&  0.018 (15)& 3 &2.579 ( 9)& 2 \cr
148.~~528   &       &                & 07 58 29& -32 18 16&  8.476 ( 6)&  0.136 ( 2)&  0.245 ( 4)&  0.989 ( 1)& 3 &2.849 (14)& 2 \cr
148.~~531   & 65680 & F3/5 IV/V      & 07 58 46& -28 39 17&  9.422 ( 4)&  0.292 ( 1)&  0.153 ( 7)&  0.509 (12)& 3 &2.670 (14)& 2 \cr
148.~~568   &       &                & 07 58 57& -31 14 19&  9.794 ( 7)&  0.323 ( 6)&  0.144 ( 0)&  0.382 (12)& 2 &2.627 ( 5)& 2 \cr
148.~~596   & 65793 & F5 V           & 07 59 20& -29 21 21&  8.417 ( 2)&  0.332 ( 2)&  0.148 ( 3)&  0.441 ( 6)& 3 &2.625 ( 1)& 2 \cr
148.~~605   &       &                & 07 59 12& -32 11 21&  9.392 ( 2)&  0.255 ( 0)&  0.162 ( 9)&  0.591 (19)& 2 &2.704 ( 5)& 2 \cr
148.~~606   & 65813 & A9 IV/V        & 07 59 28& -28 26 22&  7.670 ( 7)&  0.210 ( 0)&  0.168 ( 2)&  0.735 ( 2)& 3 &2.724 ( 7)& 2 \cr
148.~~630   &       &                & 07 59 37& -29 30 24& 10.032 ( 4)&  0.312 ( 0)&  0.160 ( 5)&  0.525 ( 7)& 2 &2.654 ( 1)& 2 \cr
148.~~635   &       &                & 07 59 33& -30 58 23&  9.761 ( 6)&  0.281 ( 1)&  0.139 ( 3)&  0.485 ( 7)& 3 &2.681 ( 8)& 2 \cr
148.~~669   &       &                & 07 59 52& -30 47 26&  9.814 (23)&  0.228 ( 7)&  0.152 (15)&  0.632 (16)& 2 &2.716 ( 2)& 2 \cr
148.~~691   & 65924 & F3 V           & 07 59 55& -31 59 26&  8.443 ( 2)&  0.292 ( 3)&  0.166 ( 1)&  0.501 ( 4)& 3 &2.668 ( 8)& 2 \cr
148.~~708   & 65959 & F5 (V)         & 08 00 15& -29 00 28&  9.124 ( 7)&  0.339 ( 2)&  0.178 ( 6)&  0.587 ( 6)& 3 &2.650 (12)& 2 \cr
148.~~710   & 65960 & F6 V           & 08 00 07& -31 46 28&  8.840 ( 2)&  0.249 ( 1)&  0.161 ( 4)&  0.642 (14)& 3 &2.708 ( 9)& 2 \cr
148.~~711   & 65961 & +F5            & 08 00 07& -31 55 28&  8.695 ( 8)&  0.353 ( 2)&  0.172 ( 5)&  0.470 ( 4)& 3 &2.645 ( 7)& 2 \cr
148.~~747   &       &                & 08 00 39& -28 59 31& 10.366 (16)&  0.154 ( 8)&  0.202 ( 9)&  0.867 ( 7)& 2 &2.820 ( 1)& 2 \cr
148.~~766   & 66074 & F0 V           & 08 00 50& -29 27 33&  8.969 ( 4)&  0.210 ( 4)&  0.192 ( 3)&  0.812 ( 3)& 2 &2.761 (13)& 2 \cr
148.~~815   & 66154 & +A2            & 08 01 04& -31 00 35& 10.033 ( 9)&  0.068 ( 0)&  0.138 ( 4)&  1.056 (24)& 2 &2.888 ( 2)& 2 \cr
148.~~860   &       &                & 08 01 18& -32 14 37&  9.928 (10)&  0.199 ( 0)&  0.214 ( 5)&  0.670 ( 9)& 2 &2.772 ( 0)& 2 \cr
148.~~887   &       &                & 08 01 38& -31 34 39&  9.575 ( 4)&  0.254 ( 2)&  0.158 ( 4)&  0.887 (19)& 2 &2.771 (29)& 2 \cr
148.~~907   & 66356 & F0 V           & 08 02 01& -29 47 42&  9.482 ( 3)&  0.259 ( 4)&  0.151 ( 1)&  0.523 ( 2)& 2 &2.689 ( 1)& 2 \cr
148.~~939   & 66399 & F3/5 V         & 08 02 12& -30 02 43&  9.836 ( 4)&  0.294 ( 1)&  0.167 ( 4)&  0.466 ( 0)& 2 &2.673 ( 7)& 2 \cr
148.~~945   & 66400 & F0 V           & 08 02 09& -31 26 43&  8.855 ( 5)&  0.239 ( 3)&  0.160 ( 4)&  0.675 ( 5)& 2 &2.708 (19)& 2 \cr
148.~~948   & 66416 & A3/5mA7-F2     & 08 02 23& -28 17 44&  9.047 ( 2)&  0.247 ( 0)&  0.252 ( 9)&  0.665 (14)& 2 &2.734 (12)& 2 \cr
148.~~965   &       &                & 08 02 22& -30 56 45& 10.195 ( 8)&  0.216 ( 1)&  0.173 ( 3)&  0.811 (20)& 2 &2.771 (10)& 2 \cr
148.~~974   &       &                & 08 02 32& -29 17 46& 10.210 ( 2)&  0.275 ( 5)&  0.132 ( 2)&  0.512 ( 0)& 2 &2.680 ( 5)& 2 \cr
148.~~995   & 66497 & A7 IV          & 08 02 45& -29 14 47&  8.576 ( 5)&  0.147 ( 2)&  0.171 ( 6)&  0.966 ( 2)& 2 &2.818 ( 8)& 2 \cr
148.1071    & 66579 & F3 V           & 08 03 15& -29 04 51&  9.557 ( 8)&  0.312 ( 2)&  0.158 ( 4)&  0.385 (11)& 2 &2.646 ( 4)& 2 \cr
148.1083    & 66621 & F0 V           & 08 03 23& -28 26 52&  9.729 ( 4)&  0.206 ( 2)&  0.171 ( 1)&  0.930 (14)& 3 &2.759 ( 5)& 2 \cr
148.1105    &       &                & 08 03 28& -28 39 53&  9.965 (11)&  0.161 ( 8)& -0.044 ( 3)& -0.059 ( 3)& 2 &2.578 (13)& 2 \cr
148.1143    & 66699 & F3/5 V         & 08 03 38& -29 37 54&  9.631 ( 3)&  0.278 ( 3)&  0.149 ( 4)&  0.466 (13)& 2 &2.681 ( 8)& 2 \cr
148.1231    &       &                & 08 04 09& -31 24 58&  9.728 (29)&  0.308 (13)& -0.086 ( 2)&  0.024 (21)& 2 &2.462 (29)& 2 \cr
148.1246    &       &                & 08 04 18& -30 11 59& 10.209 ( 8)&  0.136 ( 0)&  0.195 ( 3)&  0.851 (10)& 2 &2.825 ( 7)& 2 \cr
148.1249    &       &                & 08 04 18& -30 25 59& 10.583 ( 5)&  0.138 ( 1)&  0.174 ( 2)&  1.000 (12)& 2 &2.822 (11)& 2 \cr
148.1289    & 66966 & F5/6 V         & 08 04 46& -28 55 03&  9.625 (26)&  0.413 ( 9)&  0.234 ( 0)&  0.641 (10)& 2 &2.653 ( 1)& 2 \cr
148.1308    & 66994 & F0 IV          & 08 04 50& -29 27 03&  9.381 ( 8)&  0.168 ( 1)&  0.265 ( 2)&  0.762 ( 7)& 2 &2.821 (10)& 2 \cr
148.1347    &       &                & 08 04 57& -31 36 04& 10.114 ( 5)&  0.480 ( 2)&  0.305 (10)&  0.318 ( 6)& 2 &2.560 ( 4)& 2 \cr
148.1359    &       &                & 08 05 06& -30 28 05& 10.120 ( 2)&  0.235 ( 3)&  0.231 ( 8)&  0.808 ( 2)& 2 &2.777 ( 4)& 2 \cr
148.1371    &       &                & 08 05 16& -28 42 06& 10.585 (12)&  0.043 (17)&  0.133 (25)&  1.102 ( 8)& 2 &2.893 ( 7)& 2 \cr
148.1373    &       &                & 08 05 14& -29 30 06& 10.264 ( 3)&  0.290 ( 3)&  0.158 ( 1)&  0.504 (11)& 2 &2.686 ( 7)& 2 \cr
148.1388    & 67098 & F2 V           & 08 05 20& -29 35 07&  9.632 (14)&  0.280 ( 2)&  0.145 ( 0)&  0.533 ( 2)& 2 &2.671 ( 0)& 2 \cr
148.1418    & 67144 & F6 V(w)        & 08 05 33& -29 05 09&  9.508 ( 6)&  0.366 ( 0)&  0.166 ( 4)&  0.427 ( 8)& 2 &2.621 ( 8)& 2 \cr
148.1425    & 67146 & F2 V           & 08 05 29& -30 48 08&  8.348 ( 7)&  0.248 ( 2)&  0.150 ( 4)&  0.631 ( 1)& 2 &2.687 ( 4)& 2 \cr
148.1455    &       &                & 08 05 38& -31 55 10& 10.020 ( 7)&  0.276 ( 2)&  0.143 ( 9)&  0.498 ( 1)& 2 &2.685 ( 0)& 2 \cr
148.1476    &       &                & 08 05 56& -29 47 11& 10.468 ( 4)&  0.341 ( 9)&  0.142 ( 4)&  0.409 ( 9)& 2 &2.620 (10)& 2 \cr
148.1515    &       &                & 08 06 13& -30 03 14& 10.300 ( 3)&  0.281 ( 3)&  0.176 ( 2)&  0.641 (10)& 2 &2.700 ( 6)& 2 \cr
148.1545    &       &                & 08 06 27& -29 15 15& 10.197 ( 3)&  0.195 ( 3)&  0.167 ( 1)&  0.753 ( 1)& 2 &2.752 (16)& 2 \cr
148.1551    &       &                & 08 06 26& -29 52 15&  9.338 ( 7)&  0.246 (12)&  0.166 (12)&  0.534 ( 3)& 2 &2.712 ( 2)& 2 \cr
148.1579    &       &                & 08 06 38& -29 48 17& 10.434 ( 2)&  0.102 ( 3)&  0.171 ( 9)&  0.999 ( 8)& 2 &2.841 (21)& 2 \cr
148.1611    & 67457 & F3/5 V         & 08 06 59& -28 37 19&  9.281 ( 6)&  0.301 ( 5)&  0.154 (10)&  0.387 (13)& 2 &2.647 (37)& 2 \cr
148.1622    &       &                & 08 06 53& -30 47 19& 10.486 ( 4)&  0.345 ( 3)&  0.164 ( 3)&  0.456 ( 0)& 2 &2.638 (26)& 2 \cr
148.1650    &       &                & 08 07 16& -29 04 21& 10.780 ( 7)&  0.126 ( 6)& -0.031 ( 7)&  0.006 ( 4)& 2 &2.591 ( 3)& 2 \cr
148.1653    & 67508 & F2/3 V         & 08 07 13& -30 01 21&  9.744 ( 1)&  0.320 ( 3)&  0.123 ( 6)&  0.444 (12)& 2 &2.626 (12)& 2 \cr
148.1669    & 67554 & B2 II/III      & 08 07 29& -28 41 23&  9.716 ( 6)&  0.107 ( 7)&  0.005 ( 8)&  0.244 (11)& 2 &2.604 ( 2)& 2 \cr
148.1686    &       &                & 08 07 29& -31 00 23& 10.436 ( 8)&  0.305 ( 4)&  0.146 ( 2)&  0.484 ( 9)& 2 &2.671 ( 2)& 2 \cr
148.1707    & 67618 & F6 V           & 08 07 31& -32 11 24&  9.252 ( 3)&  0.314 ( 3)&  0.157 ( 0)&  0.401 ( 9)& 2 &2.659 ( 4)& 2 \cr
148.1755    & 67700 & F3/5 V         & 08 08 05& -30 48 28&  8.689 ( 1)&  0.280 ( 2)&  0.150 ( 6)&  0.446 ( 2)& 2 &2.661 (18)& 2 \cr
148.1842    & 67887 & F3 V           & 08 08 58& -29 14 34&  9.392 ( 7)&  0.323 ( 5)&  0.146 ( 3)&  0.492 ( 1)& 2 &2.645 ( 1)& 2 \cr
148.1883    &       &                & 08 09 04& -31 20 35& 10.471 ( 3)&  0.209 ( 2)&  0.165 ( 0)&  0.943 (15)& 2 &2.760 ( 4)& 2 \cr
148.1885    & 67922 & +A             & 08 09 04& -31 31 35&  9.393 ( 7)&  0.083 ( 1)&  0.138 ( 7)&  1.090 ( 8)& 2 &2.891 ( 2)& 2 \cr
148.1978    &       &                & 08 09 56& -29 48 41&  9.906 (12)&  0.222 ( 0)&  0.171 ( 4)&  0.709 ( 2)& 2 &2.724 (14)& 2 \cr
148.2037    & 68212 & F0/2 V         & 08 10 28& -29 21 45&  9.938 (15)&  0.171 ( 1)&  0.168 ( 5)&  0.956 ( 0)& 2 &2.811 ( 5)& 2 \cr
148.2039    & 68237 & F0 IV/V        & 08 10 27& -29 26 45&  9.082 (16)&  0.208 ( 2)&  0.177 ( 0)&  0.663 (23)& 2 &2.759 ( 4)& 2 \cr
148.2041    & 68238 & F7 V           & 08 10 25& -30 13 45&  8.431 (14)&  0.333 ( 0)&  0.163 ( 1)&  0.389 ( 2)& 2 &2.638 (17)& 2 \cr
148.2050    & 68264 & F3/5 (V)       & 08 10 35& -28 45 46&  9.349 (16)&  0.331 ( 0)&  0.169 ( 2)&  0.618 ( 4)& 2 &2.669 (26)& 2 \cr
148.2116    & 68414 & A2/3 II        & 08 11 07& -30 18 50&  9.455 (11)&  0.238 (21)&  0.267 ( 4)&  0.556 (40)& 2 &2.766 (18)& 2 \cr
148.2136    &       &                & 08 11 13& -30 21 51& 10.055 ( 7)&  0.391 ( 5)&  0.170 ( 2)&  0.430 ( 4)& 2 &2.627 ( 4)& 2 \cr
148.2164    & 68471 & F3 V           & 08 11 26& -30 01 52&  9.739 ( 8)&  0.282 ( 0)&  0.149 ( 4)&  0.478 (12)& 2 &2.672 (12)& 2 \cr
148.2182    & 68490 & A8/9 V         & 08 11 32& -29 52 53& 10.014 (26)&  0.171 (11)&  0.174 ( 8)&  0.964 ( 2)& 2 &2.800 (19)& 2 \cr
148.2200    & 68536 & A8 V           & 08 11 44& -29 54 55&  9.478 ( 1)&  0.152 ( 4)&  0.179 (12)&  0.984 (17)& 2 &2.807 ( 9)& 2 \cr
149.~~~~~~5 &       &                & 08 59 16& -31 07 20& 10.719 ( 4)&  0.187 ( 7)&  0.193 ( 9)&  0.819 (10)& 3 &2.786 (19)& 3 \cr
149.~~~~35  &       &                & 09 00 03& -31 28 24& 10.614 ( 9)&  0.344 ( 4)&  0.142 ( 3)&  0.392 (16)& 3 &2.641 (20)& 3 \cr
149.~~~~49  &       &                & 09 00 40& -31 13 28& 10.168 ( 3)&  0.502 ( 3)&  0.303 ( 5)&  0.283 ( 7)& 2 &2.572 (20)& 3 \cr
149.~~~~91  &       &                & 09 02 03& -29 11 37&  9.823 ( 5)&  0.357 ( 4)&  0.153 ( 9)&  0.498 (15)& 3 &2.640 (14)& 3 \cr
149.~~~~94  &       &                & 09 01 59& -30 55 36& 10.326 ( 3)&  0.372 ( 1)&  0.157 ( 6)&  0.501 (13)& 3 &2.650 ( 9)& 3 \cr
149.~~~~95  &       &                & 09 01 58& -31 25 36&  9.835 ( 2)&  0.281 ( 7)&  0.152 (12)&  0.526 (14)& 3 &2.697 (17)& 3 \cr
149.~~106   &       &                & 09 02 15& -31 38 38& 10.849 ( 7)&  0.235 ( 9)&  0.026 ( 3)&  1.193 (10)& 3 &2.837 (12)& 2 \cr
149.~~117   & 77591 & A7 V           & 09 02 35& -31 01 40&  7.807 (11)&  0.164 ( 3)&  0.184 ( 6)&  0.722 ( 8)& 3 &2.772 ( 7)& 3 \cr
149.~~127   & 77644 & F0 V           & 09 02 51& -31 54 42&  8.663 ( 7)&  0.234 ( 6)&  0.158 ( 7)&  0.669 ( 8)& 3 &2.710 ( 6)& 3 \cr
149.~~131   &       &                & 09 02 57& -31 47 42& 10.708 ( 7)&  0.119 ( 3)&  0.151 ( 5)&  1.065 (13)& 3 &2.874 ( 1)& 2 \cr
149.~~142   & 77715 & A9 V           & 09 03 17& -31 06 44&  8.933 ( 8)&  0.144 ( 7)&  0.207 ( 2)&  1.041 ( 2)& 3 &2.837 (13)& 3 \cr
149.~~143   &       &                & 09 03 16& -31 24 44&  9.911 ( 8)&  0.353 ( 6)&  0.156 (10)&  0.402 (11)& 3 &2.636 (17)& 2 \cr
149.~~149   &       &                & 09 03 30& -30 29 46& 10.742 ( 2)&  0.112 ( 2)&  0.165 ( 8)&  1.008 ( 5)& 2 &2.865 ( 6)& 2 \cr
149.~~161   &       &                & 09 03 46& -31 27 47& 10.462 ( 9)&  0.353 ( 3)&  0.155 ( 2)&  0.422 (16)& 2 &2.640 ( 5)& 2 \cr
149.~~164   &       &                & 09 03 55& -30 00 48& 10.403 ( 9)&  0.222 ( 1)&  0.161 ( 3)&  0.851 ( 2)& 2 &2.788 ( 3)& 2 \cr
149.~~178   &       &                & 09 04 05& -30 57 49& 10.689 ( 4)&  0.317 ( 0)&  0.139 ( 2)&  0.510 ( 9)& 2 &2.663 (12)& 2 \cr
149.~~185   & 77862 & F3/5 V         & 09 04 08& -32 10 50&  9.307 ( 2)&  0.334 ( 1)&  0.139 ( 1)&  0.453 ( 3)& 3 &2.634 ( 3)& 3 \cr
149.~~198   &       &                & 09 04 23& -30 56 51& 10.698 ( 2)&  0.097 ( 5)&  0.176 ( 5)&  1.003 (14)& 2 &2.889 (14)& 2 \cr
149.~~217   &       &                & 09 04 49& -30 04 54& 10.096 (14)&  0.277 ( 7)&  0.158 ( 9)&  0.724 ( 0)& 2 &2.697 ( 4)& 2 \cr
149.~~223   &       &                & 09 04 51& -32 10 54& 10.059 ( 0)&  0.293 ( 2)&  0.161 ( 0)&  0.577 ( 4)& 2 &2.691 ( 8)& 2 \cr
149.~~228   &       &                & 09 05 03& -32 04 55& 10.130 ( 1)&  0.269 ( 0)&  0.151 ( 1)&  0.559 ( 1)& 2 &2.705 ( 4)& 2 \cr
149.~~237   &       &                & 09 05 16& -31 31 57& 10.668 (37)&  0.217 ( 8)&  0.195 ( 9)&  0.742 (32)& 2 &2.740 ( 3)& 2 \cr
149.~~249   &       &                & 09 05 38& -29 59 59& 10.587 ( 4)&  0.217 ( 1)&  0.145 ( 0)&  0.935 (11)& 2 &2.788 (26)& 2 \cr
149.~~252   &       &                & 09 05 32& -32 20 58& 10.134 ( 4)&  0.360 ( 4)&  0.157 ( 0)&  0.433 ( 9)& 2 &2.642 ( 2)& 2 \cr
149.~~254   &       &                & 09 05 44& -29 52 59& 10.029 ( 0)&  0.352 (10)&  0.161 (13)&  0.369 (10)& 2 &2.621 (31)& 2 \cr
149.~~256   &       &                & 09 05 51& -29 12 00& 10.198 ( 2)&  0.262 ( 3)&  0.146 ( 2)&  0.670 ( 4)& 2 &2.720 (14)& 2 \cr
149.~~261   &       &                & 09 05 46& -31 34 00& 10.200 ( 2)&  0.302 ( 7)&  0.148 ( 0)&  0.627 ( 7)& 2 &2.650 ( 5)& 2 \cr
149.~~272   &       &                & 09 06 09& -29 30 02&  9.540 ( 1)&  0.421 ( 0)&  0.180 ( 2)&  0.722 ( 8)& 2 &2.673 ( 4)& 2 \cr
149.~~304   &       &                & 09 06 58& -31 49 07& 10.271 ( 4)&  0.264 ( 4)&  0.141 ( 7)&  0.495 (16)& 2 &2.702 (37)& 2 \cr
149.~~324   &       &                & 09 07 33& -29 38 10& 10.704 ( 9)&  0.115 ( 2)&  0.170 (11)&  1.047 ( 5)& 2 &2.896 (30)& 2 \cr
149.~~350   &       &                & 09 08 04& -29 02 13&  9.764 ( 1)&  0.391 ( 8)&  0.200 (14)&  0.353 (24)& 2 &2.606 ( 3)& 2 \cr
149.~~351   &       &                & 09 08 03& -29 41 13& 10.029 ( 4)&  0.265 ( 7)&  0.175 ( 4)&  0.756 ( 5)& 2 &2.749 (19)& 2 \cr
149.~~363   & 78592 & F3 V           & 09 08 12& -31 06 14&  9.421 (21)&  0.216 ( 6)&  0.221 ( 0)&  0.800 (10)& 2 &2.760 (24)& 2 \cr
149.~~373   &       &                & 09 08 30& -31 14 16& 10.823 (43)&  0.039 ( 1)&  0.121 ( 0)&  0.623 ( 4)& 2 &2.771 (10)& 2 \cr
149.~~386   &       &                & 09 08 51& -29 36 18& 10.392 ( 9)&  0.265 ( 5)&  0.175 ( 7)&  0.798 ( 0)& 2 &2.736 ( 4)& 2 \cr
149.~~391   & 78719 & F5 V           & 09 08 53& -31 47 18&  9.407 ( 9)&  0.332 ( 2)&  0.126 ( 1)&  0.343 ( 6)& 2 &2.624 ( 7)& 3 \cr
149.~~392   &       &                & 09 09 05& -28 36 19&  9.649 ( 9)&  0.402 ( 2)&  0.166 ( 2)&  0.449 ( 4)& 2 &2.621 (11)& 2 \cr
149.~~403   &       &                & 09 09 04& -32 08 20& 10.482 ( 9)&  0.334 ( 2)&  0.149 ( 0)&  0.461 ( 4)& 2 &2.662 ( 4)& 2 \cr
149.~~415   &       &                & 09 09 15& -32 19 21& 10.866 ( 9)&  0.304 ( 6)&  0.146 ( 2)&  0.585 ( 9)& 2 &2.666 (22)& 2 \cr
149.~~416   & 78858 & A5/7 II        & 09 09 26& -30 06 22& 10.692 ( 0)&  0.118 ( 9)&  0.201 (10)&  1.043 ( 0)& 2 &2.896 ( 7)& 2 \cr
149.~~420   &       &                & 09 09 40& -29 26 23& 10.018 ( 9)&  0.344 ( 0)&  0.144 ( 4)&  0.581 ( 3)& 2 &2.657 ( 6)& 2 \cr
149.~~422   &       &                & 09 09 39& -29 53 23& 10.590 ( 4)&  0.236 ( 2)&  0.160 ( 1)&  0.712 (16)& 2 &2.760 ( 2)& 2 \cr
149.~~426   &       &                & 09 09 34& -32 01 23& 11.091 ( 6)&  0.092 ( 5)&  0.159 ( 2)&  1.004 (20)& 2 &2.874 (49)& 2 \cr
149.~~437   & 78922 & A4 IV          & 09 09 56& -30 21 25&  5.629 (28)&  0.087 ( 0)&  0.202 ( 4)&  0.972 ( 6)& 2 &2.853 ( 6)& 3 \cr
149.~~460   &       &                & 09 10 28& -32 14 28& 10.006 ( 9)&  0.273 ( 2)&  0.148 ( 1)&  0.624 (11)& 2 &2.688 ( 0)& 2 \cr
149.~~466   &       &                & 09 10 43& -30 48 29& 10.128 ( 7)&  0.415 ( 4)&  0.164 ( 2)&  0.493 ( 4)& 2 &2.629 (10)& 4 \cr
149.~~469   &       &                & 09 10 41& -31 57 29& 10.753 ( 6)&  0.161 ( 9)&  0.192 (12)&  0.888 (19)& 2 &2.814 ( 8)& 2 \cr
149.~~472   &       &                & 09 10 50& -30 16 30& 10.523 (14)&  0.302 (16)&  0.158 (22)&  0.471 ( 6)& 2 &2.688 (14)& 2 \cr
149.~~482   &       &                & 09 11 09& -30 09 32&  9.809 (10)&  0.332 ( 0)&  0.133 ( 2)&  0.453 ( 0)& 2 &2.646 (10)& 3 \cr
149.~~511   &       &                & 09 11 41& -31 42 35& 10.779 (39)&  0.163 ( 4)&  0.200 ( 4)&  0.841 (16)& 2 &2.803 (31)& 2 \cr
149.~~513   &       &                & 09 11 47& -31 43 36& 10.475 (48)&  0.263 ( 2)&  0.160 ( 2)&  0.703 ( 4)& 2 &2.676 (34)& 2 \cr
149.~~520   &       &                & 09 11 55& -30 52 36& 10.093 ( 4)&  0.332 ( 2)&  0.153 ( 9)&  0.497 (14)& 2 &2.668 (13)& 3 \cr
149.~~531   &       &                & 09 12 12& -31 20 38& 10.711 ( 1)&  0.318 ( 0)&  0.154 ( 4)&  0.494 (12)& 2 &2.697 ( 3)& 2 \cr
149.~~570   &       &                & 09 12 59& -32 04 43& 10.583 ( 7)&  0.129 ( 7)&  0.151 (14)&  1.137 ( 9)& 2 &2.864 ( 7)& 2 \cr
149.~~571   &       &                & 09 12 59& -32 12 43& 10.541 ( 4)&  0.285 ( 0)&  0.150 ( 8)&  0.532 (13)& 2 &2.691 (24)& 2 \cr
149.~~583   &       &                & 09 13 22& -29 59 45& 10.398 ( 2)&  0.200 ( 1)&  0.161 ( 2)&  0.788 ( 7)& 2 &2.771 ( 2)& 2 \cr
149.~~602   &       &                & 09 13 50& -30 35 48& 10.646 ( 5)&  0.264 ( 2)&  0.157 ( 1)&  0.601 ( 7)& 2 &2.710 ( 5)& 2 \cr
149.~~614   &       &                & 09 14 08& -30 51 49&  9.921 ( 2)&  0.223 ( 2)&  0.222 ( 2)&  0.772 (11)& 2 &2.778 ( 3)& 3 \cr
149.~~630   & 79642 & A9 V           & 09 14 30& -28 40 51&  9.970 ( 2)&  0.223 ( 8)&  0.186 ( 7)&  0.672 ( 8)& 2 &2.758 ( 6)& 3 \cr
149.~~637   &       &                & 09 14 33& -30 32 52& 10.383 (12)&  0.371 (10)&  0.164 (12)&  0.403 ( 0)& 2 &2.651 ( 2)& 2 \cr
149.~~654   &       &                & 09 15 00& -29 12 54& 10.108 (20)&  0.251 ( 3)&  0.249 ( 5)&  0.679 (16)& 2 &2.786 (22)& 2 \cr
149.~~665   &       &                & 09 15 13& -31 28 56& 10.716 ( 1)&  0.335 (16)&  0.144 (17)&  0.452 (24)& 2 &2.639 ( 4)& 2 \cr
149.~~676   &       &                & 09 15 42& -29 17 58& 10.455 ( 4)&  0.298 ( 9)&  0.156 ( 4)&  0.508 (12)& 2 &2.710 ( 0)& 2 \cr
149.~~688   &       &                & 09 15 48& -32 06 59& 10.214 (16)&  0.222 ( 0)&  0.181 ( 1)&  0.700 ( 1)& 2 &2.767 ( 9)& 3 \cr
149.~~694   &       &                & 09 15 54& -32 00 00& 10.620 (12)&  0.302 ( 6)&  0.145 ( 0)&  0.469 (17)& 2 &2.688 (31)& 2 \cr
149.~~697   & 79936 & F2 V           & 09 16 01& -31 31 00&  9.611 ( 9)&  0.259 ( 4)&  0.183 ( 4)&  0.762 (13)& 2 &2.710 ( 6)& 3 \cr
149.~~712   &       &                & 09 16 30& -29 00 03& 10.149 (20)&  0.336 ( 0)&  0.157 ( 0)&  0.483 (17)& 2 &2.675 (38)& 2 \cr
149.~~722   & 80069 & A5 V           & 09 16 50& -28 26 05&  8.462 (17)&  0.171 ( 6)&  0.173 ( 4)&  0.996 ( 7)& 2 &2.791 (11)& 4 \cr
149.~~730   & 80085 & F6 V           & 09 16 51& -30 30 05&  9.401 (23)&  0.343 ( 6)&  0.140 (11)&  0.521 ( 8)& 2 &2.652 ( 5)& 3 \cr
149.~~745   &       &                & 09 17 16& -30 04 08& 10.741 (19)&  0.102 ( 0)&  0.169 ( 3)&  1.026 (22)& 2 &2.865 ( 7)& 2 \cr
149.~~748   &       &                & 09 17 28& -30 07 09&  9.915 (24)&  0.338 ( 0)&  0.172 ( 8)&  0.379 ( 9)& 2 &2.630 (10)& 2 \cr
149.~~762   &       &                & 09 17 44& -31 28 10& 10.076 (19)&  0.175 (11)&  0.179 ( 8)&  0.869 ( 6)& 2 &2.796 (16)& 3 \cr
170.~~~~~~6 & 46211 & F6/7 V         & 06 29 34& -46 48 05&  9.362 ( 5)&  0.349 ( 2)&  0.184 ( 5)&  0.452 ( 7)& 2 &2.637 ( 2)& 2 \cr
170.~~~~12  & 46235 & F2 IV/V        & 06 29 49& -46 04 07&  9.771 (12)&  0.225 ( 8)&  0.215 ( 8)&  0.708 (10)& 2 &2.732 ( 9)& 2 \cr
170.~~~~40  &       &                & 06 30 29& -46 32 13& 10.141 ( 2)&  0.294 ( 3)&  0.168 ( 4)&  0.557 ( 2)& 2 &2.663 (12)& 2 \cr
170.~~~~64  & 46433 & F5 IV          & 06 31 03& -45 49 18&  8.566 ( 4)&  0.367 ( 1)&  0.172 ( 2)&  0.504 (10)& 2 &2.620 ( 6)& 2 \cr
170.~~~~72  & 46477 & F3/5 V         & 06 31 25& -44 52 21&  9.104 ( 1)&  0.282 ( 8)&  0.156 (11)&  0.435 ( 7)& 2 &2.659 ( 2)& 2 \cr
170.~~~~81  & 46504 & A5/7 V         & 06 31 30& -46 26 22&  8.900 ( 5)&  0.140 ( 5)&  0.176 ( 5)&  0.845 ( 2)& 2 &2.779 (12)& 2 \cr
170.~~~~95  & 46626 & F2 V           & 06 32 10& -45 31 28&  9.453 ( 2)&  0.244 ( 4)&  0.159 ( 6)&  0.508 (12)& 2 &2.704 (48)& 2 \cr
170.~~107   & 46695 & F3/5 V         & 06 32 21& -46 58 30&  9.556 ( 4)&  0.305 ( 4)&  0.166 ( 8)&  0.419 (14)& 2 &2.655 (26)& 2 \cr
170.~~121   & 46743 & F5 V           & 06 32 52& -44 19 33&  8.877 ( 4)&  0.302 ( 2)&  0.151 ( 6)&  0.429 ( 5)& 2 &2.656 ( 1)& 2 \cr
170.~~149   & 46876 & F3/5 V         & 06 33 38& -43 17 40&  8.806 ( 4)&  0.284 ( 4)&  0.151 ( 5)&  0.524 ( 2)& 2 &2.665 (10)& 2 \cr
170.~~156   &       &                & 06 33 36& -45 14 40&  9.993 ( 1)&  0.330 ( 1)&  0.148 ( 1)&  0.371 (14)& 2 &2.631 (30)& 2 \cr
170.~~179   & 46998 & F0 V           & 06 34 00& -46 29 44&  9.454 (17)&  0.226 ( 3)&  0.159 ( 2)&  0.655 ( 5)& 2 &2.721 (18)& 2 \cr
170.~~200   & 47119 & F0 V           & 06 34 42& -46 29 50&  8.582 ( 2)&  0.180 ( 4)&  0.163 ( 7)&  0.790 ( 2)& 2 &2.769 ( 9)& 2 \cr
170.~~202   & 47147 & A2/5w          & 06 34 53& -45 17 51&  9.117 (31)&  0.297 (16)&  0.020 (14)&  0.802 (28)& 2 &2.623 (10)& 2 \cr
170.~~204   & 47148 & A7 V           & 06 34 47& -46 37 51&  9.683 ( 1)&  0.150 ( 0)&  0.185 ( 6)&  0.851 ( 5)& 2 &2.817 ( 6)& 2 \cr
170.~~214   &       &                & 06 35 13& -44 56 54& 10.023 ( 2)&  0.205 ( 4)&  0.136 ( 8)&  0.766 ( 4)& 2 &2.730 ( 2)& 2 \cr
170.~~243   &       &                & 06 36 15& -44 28 03&  9.737 ( 4)&  0.252 ( 3)&  0.173 ( 8)&  0.550 ( 7)& 2 &2.716 (17)& 2 \cr
170.~~247   & 47426 & A7 V           & 06 36 15& -45 47 03&  9.529 ( 0)&  0.120 ( 1)&  0.187 ( 1)&  1.082 ( 4)& 2 &2.809 ( 0)& 2 \cr
170.~~249   &       &                & 06 36 12& -46 26 03& 10.114 ( 2)&  0.358 (12)&  0.167 (19)&  0.352 ( 0)& 2 &2.628 ( 0)& 2 \cr
170.~~264   & 47501 & F2 V           & 06 36 36& -46 27 06&  8.913 ( 9)&  0.262 ( 8)&  0.153 (10)&  0.472 ( 2)& 2 &2.670 ( 4)& 2 \cr
170.~~283   &       &                & 06 37 22& -44 19 12& 10.030 ( 4)&  0.286 ( 6)&  0.132 ( 5)&  0.501 ( 2)& 2 &2.670 ( 4)& 2 \cr
170.~~306   & 47720 & F5 V           & 06 37 58& -44 17 17&  8.463 ( 3)&  0.299 ( 2)&  0.164 (10)&  0.428 (14)& 2 &2.651 (16)& 2 \cr
170.~~320   &       &                & 06 38 24& -43 50 21& 10.308 ( 2)&  0.197 ( 0)&  0.171 ( 1)&  0.738 ( 0)& 2 &2.751 ( 4)& 2 \cr
170.~~336   &       &                & 06 38 51& -43 20 25&  9.628 ( 1)&  0.227 ( 0)&  0.164 ( 4)&  0.615 (12)& 2 &2.716 ( 6)& 2 \cr
170.~~365   &       &                & 06 39 39& -43 21 31&  9.006 ( 7)&  0.267 ( 1)&  0.154 ( 4)&  0.475 ( 6)& 2 &2.671 (15)& 2 \cr
170.~~382   & 48168 & F3/5 V         & 06 39 42& -46 38 33&  9.659 ( 4)&  0.289 ( 4)&  0.158 ( 5)&  0.482 ( 4)& 2 &2.662 (15)& 2 \cr
170.~~390   & 48243 & A7 V           & 06 40 13& -43 41 37&  8.570 ( 4)&  0.150 ( 7)&  0.185 ( 9)&  0.800 ( 3)& 2 &2.798 ( 2)& 2 \cr
170.~~418   & 48384 & F3/5 V         & 06 40 52& -45 44 43&  9.655 ( 2)&  0.297 ( 3)&  0.158 ( 0)&  0.492 ( 2)& 2 &2.648 (24)& 2 \cr
170.~~431   & 48429 & A7 V           & 06 41 13& -45 06 46&  8.668 (14)&  0.160 ( 2)&  0.185 ( 1)&  0.867 ( 4)& 2 &2.787 ( 8)& 2 \cr
170.~~436   & 48464 & F0 V           & 06 41 18& -45 16 46&  8.989 ( 1)&  0.144 ( 2)&  0.208 ( 9)&  0.755 ( 9)& 2 &2.798 ( 5)& 2 \cr
170.~~483   & 48730 & F3/5 IV        & 06 42 30& -45 20 57&  9.535 ( 0)&  0.297 ( 0)&  0.179 ( 7)&  0.619 (12)& 2 &2.697 ( 7)& 2 \cr
170.~~488   & 48744 & F2 V           & 06 42 28& -47 01 57&  9.438 ( 0)&  0.229 (10)&  0.168 (13)&  0.573 (11)& 2 &2.711 (12)& 2 \cr
170.~~492   &       &                & 06 42 57& -43 21 00&  9.540 ( 2)&  0.306 ( 8)&  0.132 (14)&  0.381 ( 3)& 2 &2.641 ( 0)& 2 \cr
170.~~516   & 48856 & A8 III         & 06 43 04& -47 08 02&  9.345 ( 2)&  0.178 ( 2)&  0.169 ( 0)&  0.707 ( 1)& 2 &2.761 ( 2)& 2 \cr
170.~~524   & 48905 & A3 III         & 06 43 36& -44 03 06&  8.274 ( 3)&  0.082 ( 1)&  0.204 ( 2)&  0.974 ( 1)& 2 &2.865 (24)& 2 \cr
170.~~564   &       &                & 06 44 46& -44 33 16&  9.974 ( 2)&  0.332 ( 0)&  0.178 ( 2)&  0.397 (11)& 2 &2.659 (12)& 2 \cr
170.~~568   &       &                & 06 44 39& -46 07 15& 10.297 ( 1)&  0.233 ( 7)&  0.259 (12)&  0.742 (14)& 2 &2.758 ( 0)& 2 \cr
170.~~573   &       &                & 06 45 01& -43 56 18&  9.991 ( 6)&  0.219 ( 8)&  0.178 ( 4)&  0.657 (19)& 2 &2.723 (13)& 2 \cr
170.~~580   & 49216 & F3 IV/V        & 06 45 13& -43 52 20&  9.015 ( 1)&  0.286 ( 4)&  0.166 ( 6)&  0.591 (14)& 2 &2.681 (13)& 2 \cr
170.~~583   & 49217 & F0 IV/V        & 06 45 02& -46 19 19&  9.444 (12)&  0.176 ( 0)&  0.172 ( 1)&  0.702 ( 2)& 2 &2.763 (10)& 2 \cr
170.~~586   &       &                & 06 45 16& -44 27 20&  9.729 ( 7)&  0.306 ( 8)&  0.162 (12)&  0.401 ( 9)& 2 &2.643 ( 3)& 2 \cr
170.~~603   &       &                & 06 45 40& -44 31 24&  9.766 (11)&  0.319 ( 7)&  0.152 ( 8)&  0.390 ( 7)& 2 &2.626 (15)& 2 \cr
170.~~614   & 49394 & F2 V           & 06 46 07& -43 48 27&  8.829 ( 4)&  0.248 ( 5)&  0.152 ( 4)&  0.498 ( 4)& 2 &2.699 (21)& 2 \cr
170.~~618   & 49421 & F2 IV/V        & 06 46 16& -44 36 29&  9.092 ( 2)&  0.213 ( 5)&  0.170 ( 9)&  0.782 ( 2)& 2 &2.742 ( 4)& 2 \cr
170.~~642   & 49577 & F5 V           & 06 46 58& -44 41 35&  9.105 ( 3)&  0.321 ( 2)&  0.141 ( 7)&  0.425 (10)& 2 &2.624 ( 5)& 2 \cr
170.~~665   & 49755 & A3/7           & 06 47 38& -46 23 41&  9.998 ( 4)&  0.133 ( 4)&  0.170 ( 2)&  1.060 ( 1)& 2 &2.813 (21)& 2 \cr
170.~~666   & 49756 & F3 V           & 06 47 37& -46 33 41&  9.777 ( 3)&  0.275 ( 0)&  0.153 ( 3)&  0.466 (28)& 2 &2.667 (19)& 2 \cr
170.~~688   & 49919 & F2 IV          & 06 48 38& -43 37 49&  8.634 ( 1)&  0.232 ( 5)&  0.166 ( 7)&  0.661 (12)& 2 &2.695 ( 2)& 2 \cr
170.~~697   & 49963 & F3 IV/V        & 06 48 41& -45 44 50&  9.369 ( 8)&  0.278 ( 1)&  0.169 ( 0)&  0.457 ( 7)& 2 &2.672 ( 9)& 2 \cr
170.~~698   &       &                & 06 48 59& -43 08 52&  9.572 ( 5)&  0.318 ( 8)&  0.172 (10)&  0.397 ( 9)& 2 &2.641 (25)& 2 \cr
170.~~710   &       &                & 06 49 07& -46 34 54&  9.710 ( 6)&  0.337 ( 2)&  0.149 ( 1)&  0.497 (10)& 2 &2.657 ( 4)& 2 \cr
170.~~723   & 50181 & A2+F/G (III)   & 06 49 50& -45 02 59&  8.290 ( 7)&  0.243 ( 0)&  0.186 ( 1)&  0.675 ( 5)& 2 &2.726 (12)& 2 \cr
170.~~724   &       &                & 06 50 01& -44 02 01& 10.340 (14)&  0.180 (10)&  0.170 (14)&  0.718 ( 2)& 2 &2.756 ( 3)& 2 \cr
170.~~741   & 50308 & F3/5 V         & 06 50 20& -46 22 04&  9.605 ( 9)&  0.285 ( 0)&  0.147 ( 0)&  0.433 ( 6)& 2 &2.664 ( 9)& 2 \cr
170.~~744   & 50334 & F2 V           & 06 50 27& -46 06 05&  9.589 (11)&  0.263 ( 2)&  0.139 ( 2)&  0.508 ( 7)& 2 &2.676 ( 0)& 2 \cr
170.~~745   & 50361 & F0 IV/V        & 06 50 44& -43 41 07&  8.802 ( 9)&  0.214 ( 0)&  0.165 ( 6)&  0.665 (16)& 2 &2.720 ( 2)& 2 \cr
170.~~752   &       &                & 06 50 35& -47 07 07&  9.892 (17)&  0.296 ( 4)&  0.152 ( 2)&  0.493 (13)& 2 &2.669 ( 2)& 2 \cr
171.~~~~57  & 59970 & F3/5 V         & 07 30 20& -43 59 37&  9.230 ( 0)&  0.312 ( 7)&  0.150 ( 2)&  0.466 ( 1)& 2 &2.657 ( 0)& 2 \cr
171.~~~~59  &       &                & 07 30 08& -46 40 36&  9.456 ( 0)&  0.367 (10)&  0.175 ( 9)&  0.392 (11)& 2 &2.619 ( 9)& 2 \cr
171.~~~~68  & 60032 & F3 V           & 07 30 31& -45 39 39&  8.421 ( 4)&  0.312 ( 9)&  0.134 ( 6)&  0.428 ( 2)& 2 &2.642 ( 9)& 2 \cr
171.~~171   &       &                & 07 33 24& -45 57 02&  9.435 (12)&  0.368 ( 0)&  0.157 ( 0)&  0.468 (10)& 2 &2.656 ( 4)& 2 \cr
171.~~485   & 62062 & F5 V           & 07 39 59& -45 09 54&  8.621 ( 5)&  0.300 ( 1)&  0.148 ( 4)&  0.471 ( 1)& 2 &2.671 (11)& 2 \cr
171.~~550   &       &                & 07 41 18& -43 37 04&  9.150 ( 5)&  0.407 ( 0)&  0.164 ( 7)&  0.497 (25)& 2 &2.650 ( 2)& 2 \cr
171.~~710   &       &                & 07 44 18& -46 42 29& 10.425 ( 3)&  0.276 ( 2)&  0.155 ( 6)&  0.555 (16)& 2 &2.710 ( 4)& 2 \cr
171.~~846   &       &                & 07 47 05& -44 14 50& 10.017 ( 4)&  0.343 ( 6)&  0.162 ( 2)&  0.401 ( 4)& 2 &2.631 (24)& 2 \cr
171.~~956   &       &                & 07 49 08& -43 37 06& 10.095 ( 6)&  0.248 ( 2)&  0.159 ( 4)&  0.565 ( 8)& 2 &2.712 (16)& 2 \cr
173.~~~~60  & 82152 & F7 V           & 09 28 48& -43 48 13&  8.111 ( 0)&  0.324 ( 1)&  0.159 ( 2)&  0.425 ( 1)& 2 &2.643 (16)& 2 \cr
173.~~~~78  &       &                & 09 29 03& -46 51 15&  9.722 ( 2)&  0.365 ( 1)&  0.164 ( 0)&  0.450 ( 9)& 2 &2.639 ( 5)& 2 \cr
173.~~126   &       &                & 09 30 21& -45 17 22&  9.294 ( 5)&  0.312 ( 0)&  0.151 ( 9)&  0.408 ( 6)& 2 &2.657 ( 6)& 2 \cr
173.~~147   &       &                & 09 30 48& -46 07 24& 10.390 ( 4)&  0.318 ( 4)&  0.162 ( 1)&  0.382 ( 7)& 2 &2.667 (14)& 2 \cr
173.~~326   &       &                & 09 34 54& -46 55 46& 10.255 ( 5)&  0.232 ( 1)&  0.173 ( 4)&  0.730 (13)& 2 &2.721 (29)& 2 \cr
173.~~338   &       &                & 09 35 12& -46 57 48& 10.828 ( 8)&  0.233 ( 2)&  0.151 (14)&  0.559 (12)& 2 &2.679 (17)& 2 \cr
173.~~384   &       &                & 09 36 18& -45 12 53&  9.300 ( 0)&  0.286 ( 4)&  0.133 ( 4)&  0.556 ( 2)& 2 &2.667 (12)& 2 \cr
173.~~389   &       &                & 09 36 18& -47 15 53&  9.807 ( 4)&  0.235 ( 4)&  0.179 ( 4)&  0.649 ( 2)& 2 &2.735 ( 5)& 2 \cr
173.~~413   &       &                & 09 36 58& -45 44 57&  9.962 ( 2)&  0.310 ( 3)&  0.172 ( 0)&  0.406 ( 7)& 2 &2.658 ( 0)& 2 \cr
173.~~444   &       &                & 09 37 33& -46 27 00&  9.860 ( 1)&  0.288 ( 1)&  0.157 ( 0)&  0.414 ( 0)& 2 &2.652 (11)& 2 \cr
173.~~449   &       &                & 09 37 43& -45 04 00& 10.329 ( 1)&  0.312 ( 1)&  0.138 ( 2)&  0.477 ( 4)& 2 &2.657 ( 0)& 2 \cr
173.~~478   &       &                & 09 38 24& -45 24 04& 10.732 (14)&  0.270 (10)&  0.135 (15)&  0.561 (24)& 2 &2.695 ( 0)& 2 \cr
173.~~488   &       &                & 09 38 37& -45 00 05&  9.885 ( 5)&  0.312 ( 5)&  0.149 ( 9)&  0.518 (10)& 2 &2.657 (16)& 2 \cr
173.~~509   & 83707 & F3 IV          & 09 38 55& -47 19 07&  9.216 ( 4)&  0.280 ( 5)&  0.164 ( 9)&  0.565 ( 2)& 2 &2.679 ( 7)& 2 \cr
173.~~539   &       &                & 09 39 41& -46 02 11& 10.469 ( 2)&  0.496 ( 7)&  0.270 ( 4)&  0.417 ( 0)& 2 &2.607 ( 4)& 2 \cr
173.~~551   &       &                & 09 39 51& -46 48 11& 11.031 ( 1)&  0.203 ( 0)&  0.165 ( 5)&  0.816 ( 7)& 2 &2.756 (21)& 2 \cr
173.~~554   &       &                & 09 40 02& -45 00 12&  9.764 ( 0)&  0.399 ( 7)&  0.184 ( 3)&  0.396 ( 0)& 2 &2.616 (11)& 2 \cr
173.~~559   &       &                & 09 40 18& -43 39 13& 10.223 ( 1)&  0.322 ( 2)&  0.170 ( 2)&  0.572 ( 7)& 2 &2.662 (11)& 2 \cr
173.~~642   &       &                & 09 41 38& -47 27 21& 10.651 ( 0)&  0.281 ( 0)&  0.139 ( 3)&  0.754 (23)& 2 &2.713 (16)& 2 \cr
173.~~702   &       &                & 09 42 53& -46 25 27& 10.760 ( 0)&  0.341 ( 4)&  0.146 ( 2)&  0.518 (10)& 2 &2.641 (10)& 2 \cr
173.~~708   &       &                & 09 43 11& -44 21 28& 10.711 ( 2)&  0.204 ( 0)&  0.169 ( 0)&  0.721 ( 1)& 2 &2.753 (16)& 2 \cr
173.~~753   &       &                & 09 44 00& -46 17 32& 10.418 ( 6)&  0.351 ( 2)&  0.165 ( 9)&  0.427 ( 0)& 2 &2.627 (27)& 2 \cr
173.~~821   &       &                & 09 45 09& -45 31 38&  9.950 ( 4)&  0.306 ( 0)&  0.154 ( 0)&  0.418 ( 3)& 2 &2.636 ( 3)& 2 \cr
173.~~830   &       &                & 09 45 15& -45 28 38& 10.544 ( 2)&  0.179 ( 1)&  0.245 ( 2)&  0.742 (10)& 2 &2.808 (19)& 2 \cr
173.~~870   &       &                & 09 45 57& -45 31 42& 10.249 ( 2)&  0.279 ( 2)&  0.164 ( 0)&  0.665 (14)& 2 &2.696 ( 3)& 2 \cr
173.~~902   &       &                & 09 46 29& -47 23 44& 10.132 ( 5)&  0.252 ( 7)&  0.169 (14)&  0.752 (21)& 2 &2.731 (17)& 2 \cr
173.~~942   &       &                & 09 47 34& -45 42 49&  9.895 ( 2)&  0.247 ( 2)&  0.171 ( 7)&  0.651 ( 9)& 2 &2.704 ( 0)& 2 \cr
173.~~970   &       &                & 09 48 02& -46 35 52& 10.767 ( 0)&  0.317 ( 2)&  0.028 ( 2)&  0.775 ( 2)& 2 &2.742 (19)& 2 \cr
173.~~985   &       &                & 09 48 21& -46 10 53& 10.625 ( 9)&  0.219 ( 4)&  0.174 ( 7)&  0.837 (12)& 2 &2.761 (23)& 2 \cr
173.~~991   &       &                & 09 48 26& -46 26 54& 10.731 ( 2)&  0.243 ( 0)&  0.164 ( 3)&  0.760 ( 9)& 2 &2.735 (11)& 2 \cr
173.1005    &       &                & 09 48 51& -43 54 55&  9.668 ( 4)&  0.288 ( 4)&  0.165 ( 4)&  0.488 ( 5)& 2 &2.702 ( 4)& 2 \cr
173.1051    & 85312 & F2 V           & 09 49 37& -47 13 59&  9.520 ( 2)&  0.308 ( 0)&  0.148 ( 2)&  0.510 (12)& 2 &2.655 ( 5)& 3 \cr
\hline
\end{longtable}
}

\section{Data analysis}

\subsection{Observed colour excesses}
\label{obs_col_exc}

Intrinsic colours and absolute stellar magnitudes can be computed in the $uvby$$-$$\beta$ system
for stars of spectral types ranging from B to early/mid G-type stars, excluding A1 and A2 stars, 
following the calibrations suggested by \citet{CR75,CR78,CR79}. To obtain intrinsic values
with a high degree of confidence, a set of selection criteria were imposed on the observed sample.
This procedure has the purpose of avoiding objects with peculiar indices in our final sample
\citep[a detailed description of the adopted selection criteria is given by][]{FR89a}. A total of 520 
stars, from the combined samples, fulfilled the imposed selection criteria. About 96\% of them
belong to the late AF type star group.

%--------------------------- Figure 2 ----------------------------------------
\begin{figure}
\centering
\includegraphics[width=\hsize]{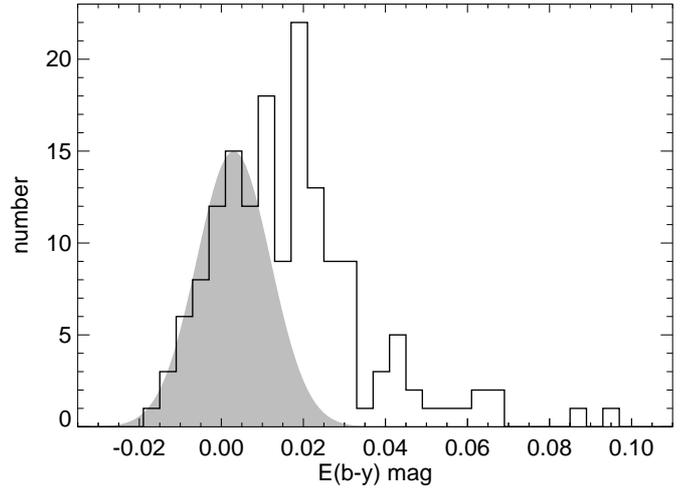}
\caption{Distribution of the obtained colour excesses for stars closer than 200\,pc (bin = $0\fm004$). 
The shadowed area represents a standard normal (Gaussian) distribution centred 
on E$(b-y) = 0\fm003$ and a standard deviation $\sigma = 0\fm009$, which is basically equal to 
the computed mean accuracy of the estimated colour excess ($\sigma_{{\rm E}(b-y)} = 0\fm010$).}
\label{histo}
\end{figure}
%------------------------------------------------------------------------------

The standard deviations of the combined sample are slightly larger than the ones obtained for the 
new data introduced in Table\,\ref{photometry}, and are $0\fm007$, $0\fm004$, $0\fm006$, 
$0\fm009$, and $0\fm011$, for $V$, $(b-y)$, $m_1$, $c_1$, and H$\beta$, respectively. The 
overall accuracy of the obtained colour excesses can be estimated by propagation of the
measurements errors into the calibrations. On the basis of the above-mentioned standard deviations,
the mean accuracy of the colour excesses was estimated to be better than $0\fm010$.

Figure\,\ref{histo} displays the obtained distribution of colour excesses for stars having estimated 
distances closer than 200\,pc from the Sun. A total of 157 stars are found within this distance limit. 
For purpose of comparison, a standard normal (Gaussian) distribution centred on E$(b-y) = 0\fm003$ 
and standard deviation $\sigma= 0\fm009$ is given by the shadowed area. The left side of the 
obtained distribution of colour excesses seems to be rather well-represented by the normal 
distribution, an indication that the estimated mean accuracy ($\sigma_{{\rm E}(b-y)} = 0\fm010$) is 
close to the true value for the investigated sample. The distribution of colour excesses clearly show 
a number of reddened stars that cannot be explained by measurement uncertainties alone, in 
particular, for the bin around E$(b-y) \sim 0\fm02$. This reddening is likely to be associated with
the absorption due to single small diffuse clouds. The reddening in a single cloud, $e_o$, and the 
number of these clouds per unit length, $\nu$, can be deduced by applying the method proposed 
by \citet{muench52}, which, in the present case, provides $e_o = 0\fm023$ and $\nu = 4.1$ clouds 
per kpc. These values are slightly smaller than the ones obtained by \citet{knude79} for the solar 
vicinity, mainly the number of clouds per kpc, which in his case is about one and a half to two times 
larger than the value found here. This is understandable because the volume probed here seems 
to be more depleted than the average solar neighbourhood.  

For a density of four clouds per kpc, it is expected that some observed lines-of-sight traverse a 
single diffuse cloud within the 200\,pc distance interval, and that a few might even traverse
two clouds, which would explain the small peak observed close to the $0\fm04$ bin. The observed 
excess of stars with reddening around E$(b-y) \sim 0\fm01$, however, remains unexplained.

\subsection{Stellar distances}

\begin{figure}
\centering
\includegraphics[width=\hsize]{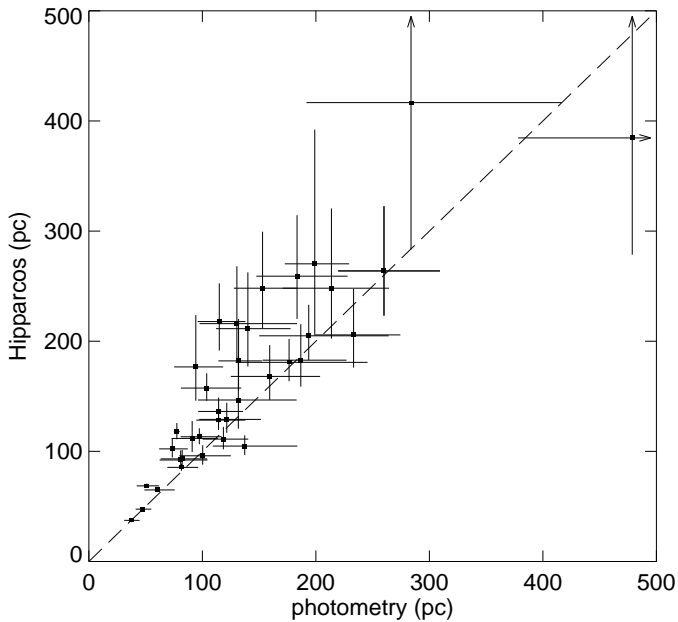}
\caption{Comparison between the trigonometric distances obtained from the new {\it Hipparcos} 
catalogue  \citep{leeuwen07} and the photometric distances based on $uvby$$-$$\beta$ data. The 
errors indicated for the photometric distances were estimated by propagating the measured
photometric uncertainties into the calibrations of each star. The diagonal dashed-line is shown 
for reference only.}
\label{hipparcos}
\end{figure}

The accuracy of the photometric distance determination depends on the stellar 
spectral type. For the A- and F-type stars, it is estimated that the distances have, in
average, an accuracy of better than 30\%, while distances for late B-type stars are better
than 20\% and for early B-type stars better than 40\%. An independent test of this accuracy 
can be obtained by comparing with parallactic distances obtained by the {\it Hipparcos} satellite.
By searching in the new {\it Hipparcos} catalogue \citep{leeuwen07} we discovered 37 stars in 
common with the photometric sample. The comparison between both estimated distances
is shown in Figure\,\ref{hipparcos}. In spite of the moderate trend for the photometric distances 
to be slightly underestimated compared to the parallactic one, the agreement for most 
stars is very good, testifying to the high quality of the photometric distance determination.

%--------------------------- Figure 3 ----------------------------------------
%-----------------------------------------------------------------------------

\subsection{Colour excess {\it versus} distance diagrams}\label{colour_dis}

\begin{figure*}
\centering
\includegraphics[width=\hsize]{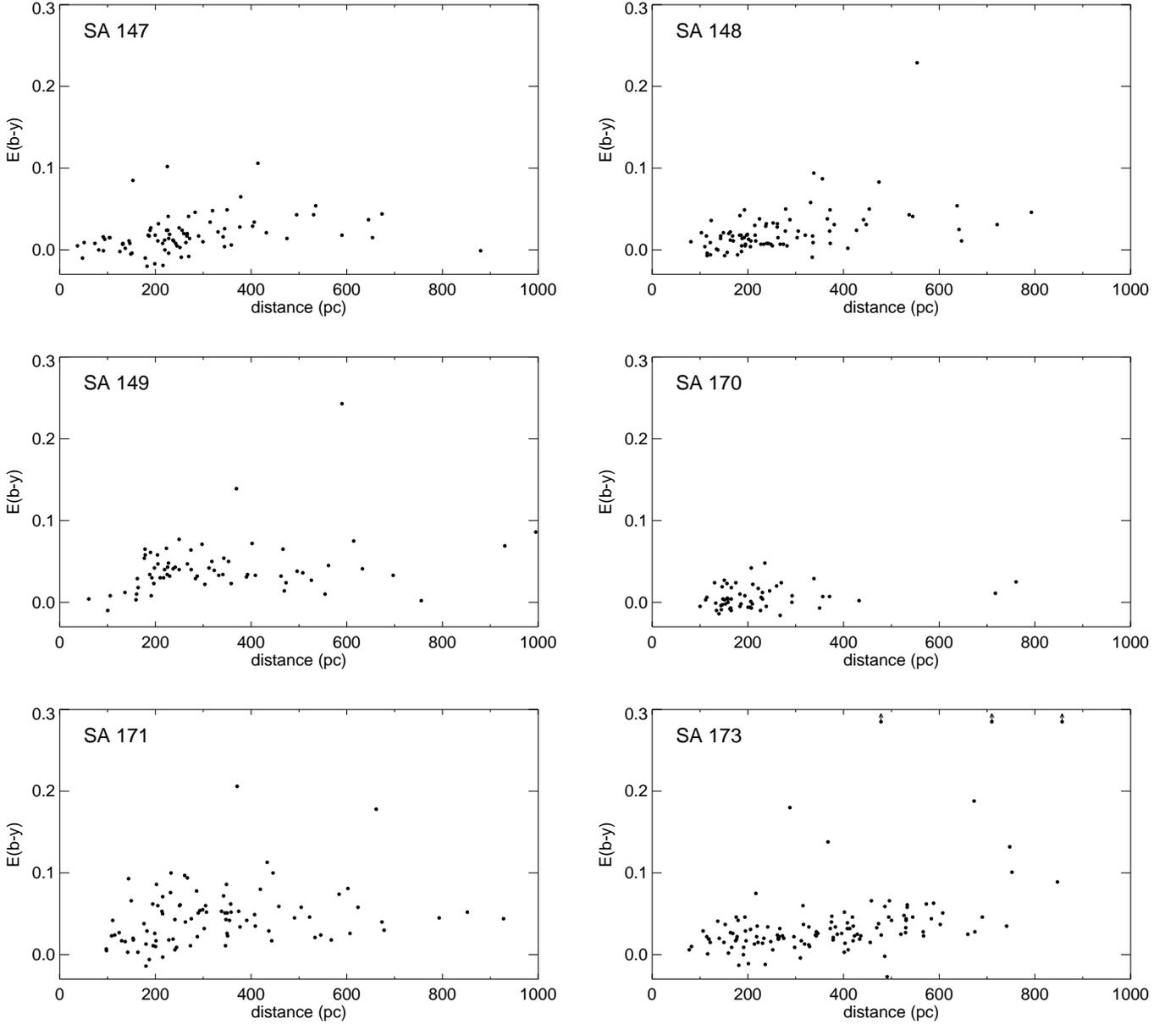}
\caption{Colour excess {\it vs.} distance diagrams obtained for each of the observed selected area. 
Three stars belonging to SA\,173 present colour excess larger than $0\fm3$ at distances 
smaller than 1\,kpc. Their distances are indicated in the diagram by dots with arrows.}
\label{excess}
\end{figure*}

The obtained colour excess {\it versus} distance diagrams for each observed area are given 
in Fig.\,\ref{excess}. A common characteristic noted for all of these areas is the rather low 
level of interstellar absorption experienced by most of the observed stars, not only for the
nearest of them, but out to about 1\,kpc from the Sun. 

As already  mentioned, two of the observed areas have a line-of-sight that is partially outside 
the assumed rim of the Gum nebula. SA\,170 seems to be the volume with the lowest 
interstellar density. The obtained colour excesses for this selected area do not exceed 
E$(b-y) = 0\fm05$ and yield a mean colour excess of $\langle {\rm E}(b-y) \rangle = 0\fm006$. 
All observed stars in SA\,170 are basically closer than 400\,pc  from the Sun. Beyond that, there 
are only two other stars with estimated distances ranging from 700\,pc to 800\,pc and colour 
excesses not exceeding $0\fm030$. The other area is SA\,147, which is located about 15\degr\ to the 
north of SA\,170. Taking into account only stars closer than 200\,pc, except for one, the obtained mean 
colour excess for SA\,147 is also $\langle {\rm E}(b-y) \rangle = 0\fm006$. The excluded star is 
SA147.1218 (HD\,55447). From the photometry, we estimated a colour excess 
E$(b-y) = 0\fm085$ and a distance of 153$\pm27$ pc \citep[the estimated parallactic distance is 
$248^{+51}_{-37}$\,pc,][]{leeuwen07}. Beyond 200\,pc, it is possible to note the effect of 
some interstellar absorption --- the colour excess as a function of the distance shows a tendency 
to increase and the lower envelope of the data points distribution gets slightly higher.

While the two former selected areas have lines-of-sight through directions of rather low 
interstellar extinction (see Fig.\,\ref{sfd_map}), SA\,148 probes a volume that is expected to
have a higher absorption. Nevertheless, the obtained colour excess $versus$ distance diagram for
this area resembles those obtained for these areas. Out to about 320\,pc, the 
derived colour excess does not exceed E$(b-y) = 0\fm05$, with a mean colour excess of 
$\langle {\rm E}(b-y) \rangle = 0\fm015$. Beyond the distance of 200\,pc, there is clearly a
slight increase in the value of the lower envelope of the data points, although, the maximum
value remains constant up to about the aforementioned distance of $\sim$320\,pc, after what, some 
absorption sets up. Beyond $\sim$400\,pc the observed minimum colour-excess seems to increase.
Although one star has a colour excess of E($b-y) \approx 0\fm23$ at a distance of 
$\sim$550\,pc, the general behaviour of the colour excess seems to continue the same, that is, 
around E$(b-y) \approx 0\fm05$ up to about 800\,pc. Our entire photometric sample contains eight
stars with estimated distances larger than 1.0\,kpc, five of then belonging to SA\,148. The 
estimated colour excess for these stars are in the range from E$(b-y) =0\fm20$ to E$(b-y) = 0\fm35$.

%--------------------------- Figure 4 ----------------------------------------
%-----------------------------------------------------------------------------

SA\,149 displays an interesting colour excess {\it versus} distance diagram. The eight closest
stars ($d < 170$\,pc) have E$(b-y) < 0\fm03$. Beyond this distance, the colour excess suffers
a steep transition, which is clearly shown in terms of both an increase in the maximum absorption 
and the value of the lower envelope. Nevertheless, the maximum colour excess does not 
exceed E$(b-y) = 0\fm09$, except in two cases  (E$(b-y) = 0\fm14$ at $d = 370$\,pc and 
E$(b-y) = 0\fm25$ at $d= 590$\,pc).

The remaining two selected areas, SA\,171 and SA\,173, were previously analysed by 
\citet{FR90}, and the colour excess {\it versus} distance diagrams presented there
are pretty similar to the ones given in the present work. Nevertheless, it is instructive to return to
these areas, first of all, because new data have been incorporated into the observational sample
(nine stars in SA\,171 and thirty-two in SA\,173). We note that the analysis conducted in the 
previous investigation was done prior to the knowledge of IVS towards the same line-of-sight.

%--------------------------- Figure 5 ----------------------------------------
\begin{figure*}
\begin{center}
\hbox{\includegraphics[width=.48\hsize]{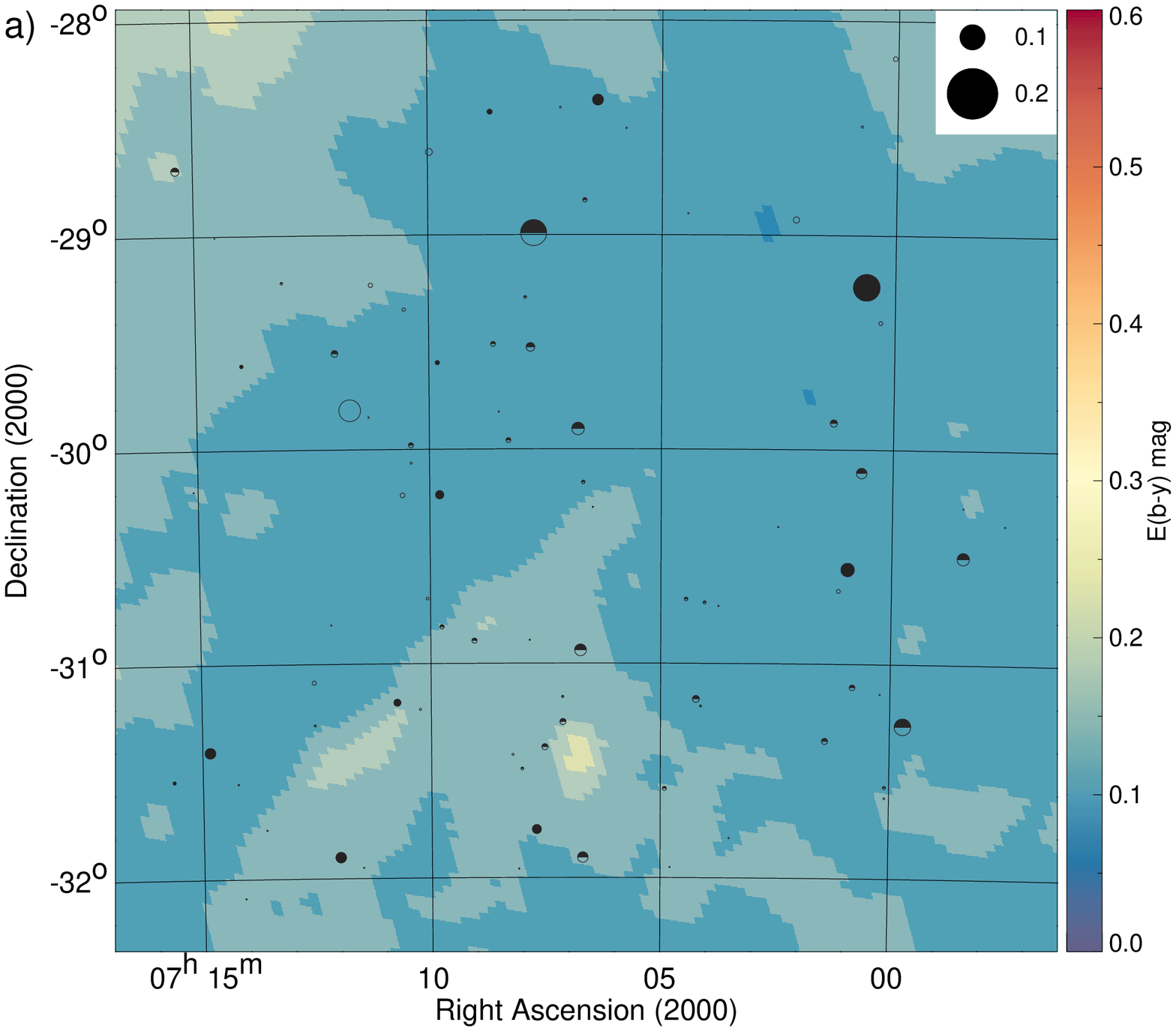}\hskip.20cm\includegraphics[width=.48\hsize]{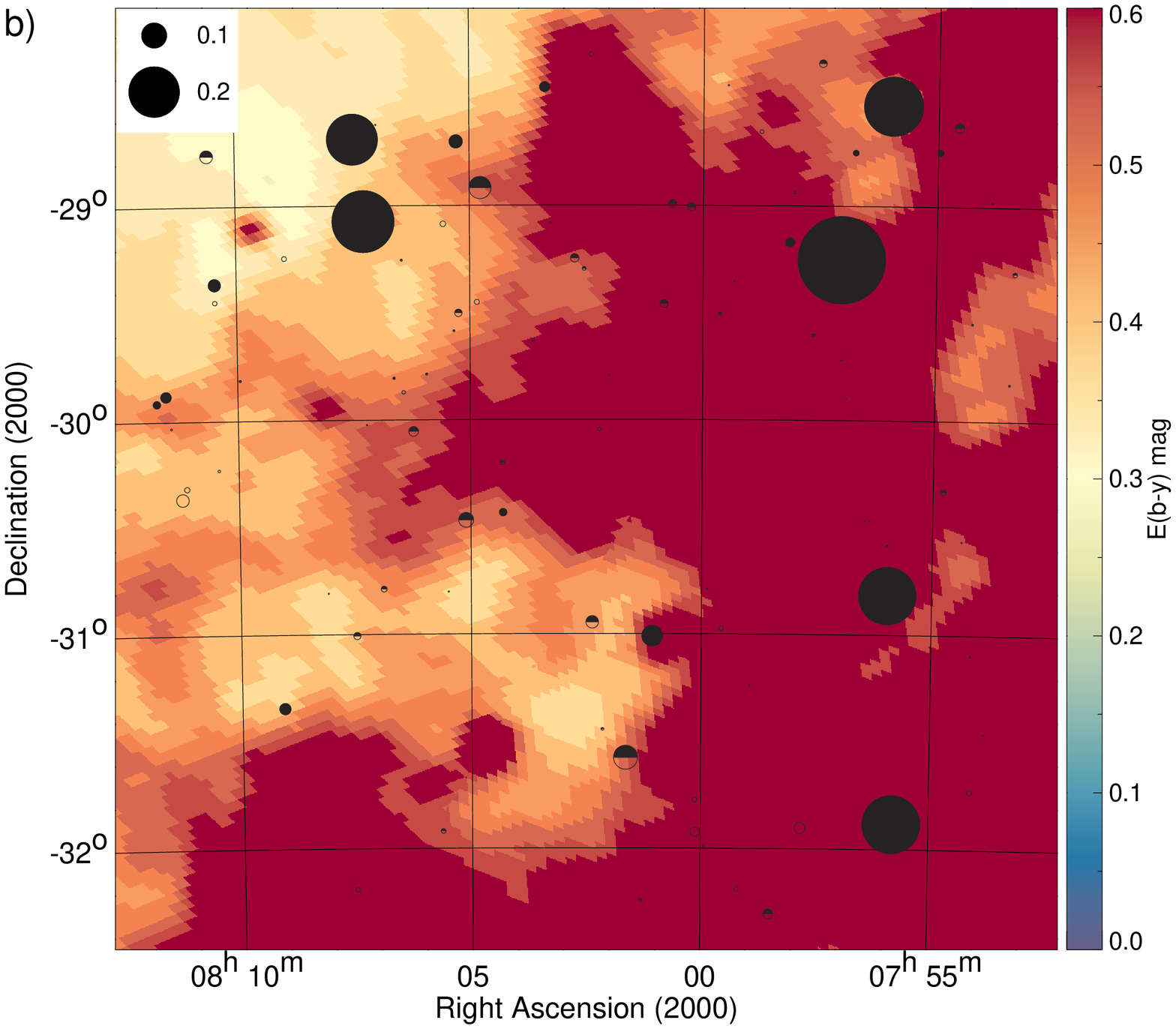}}
\hbox{\includegraphics[width=.48\hsize]{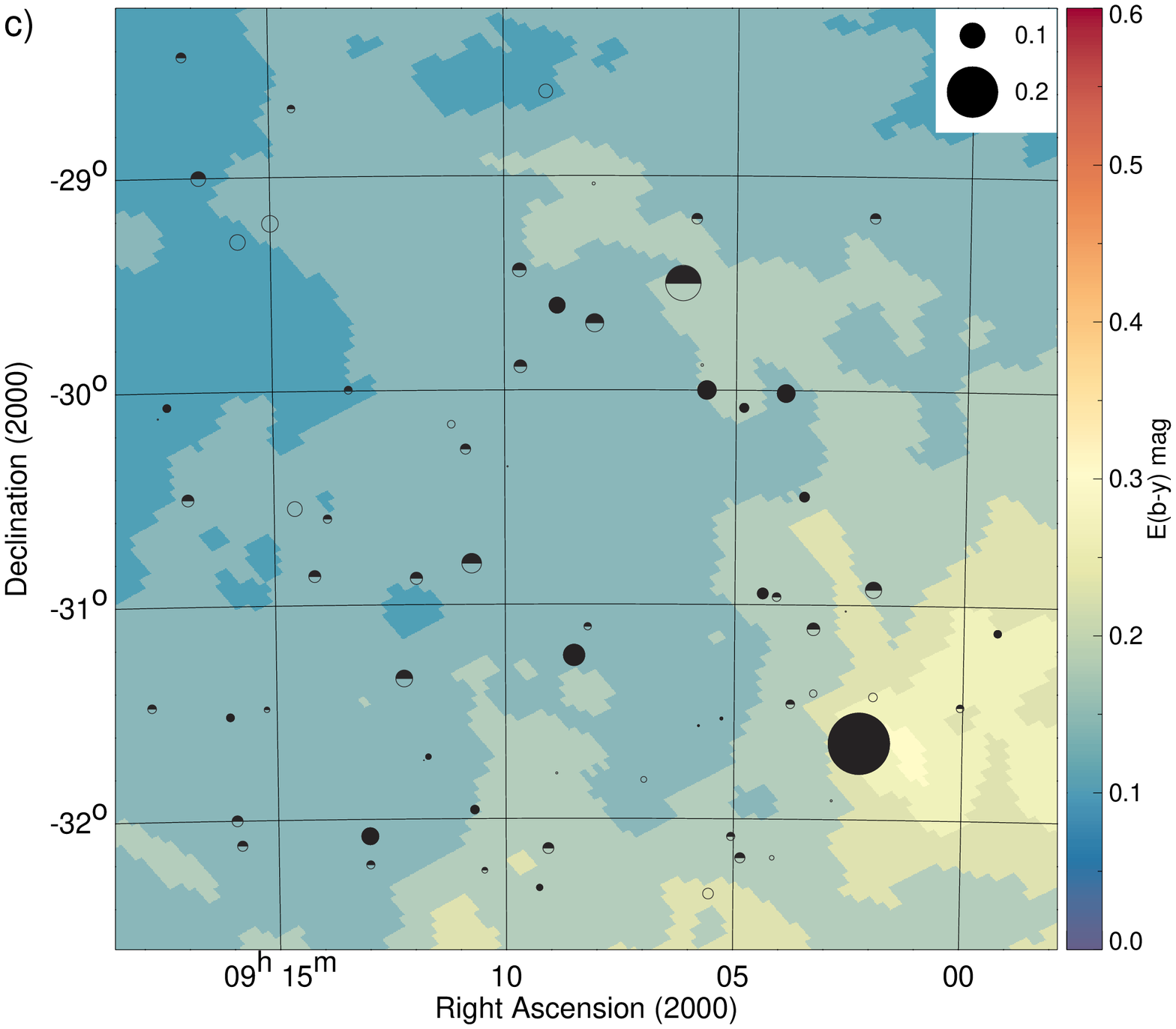}\hskip.20cm\includegraphics[width=.48\hsize]{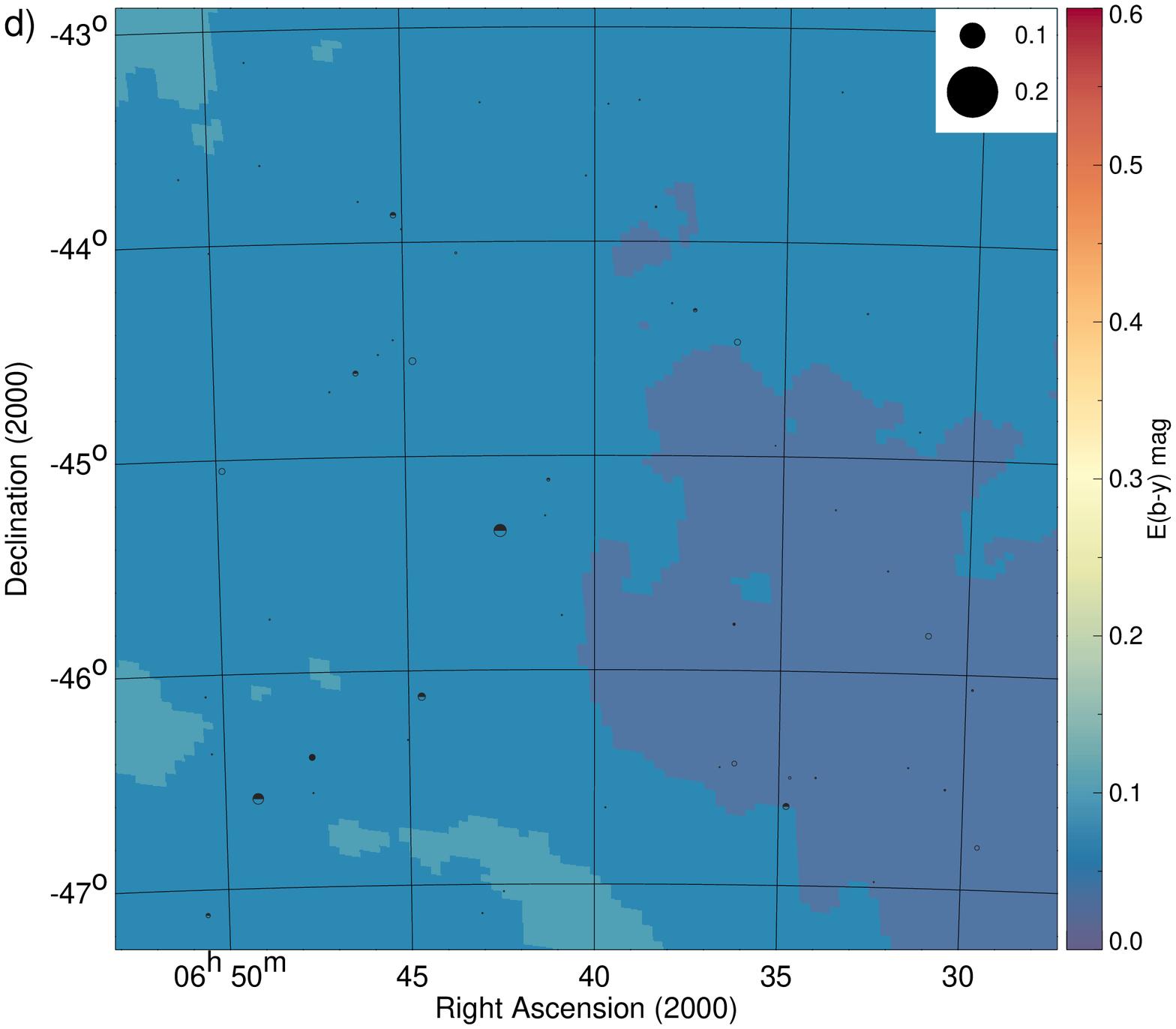}}
\hbox{\includegraphics[width=.48\hsize]{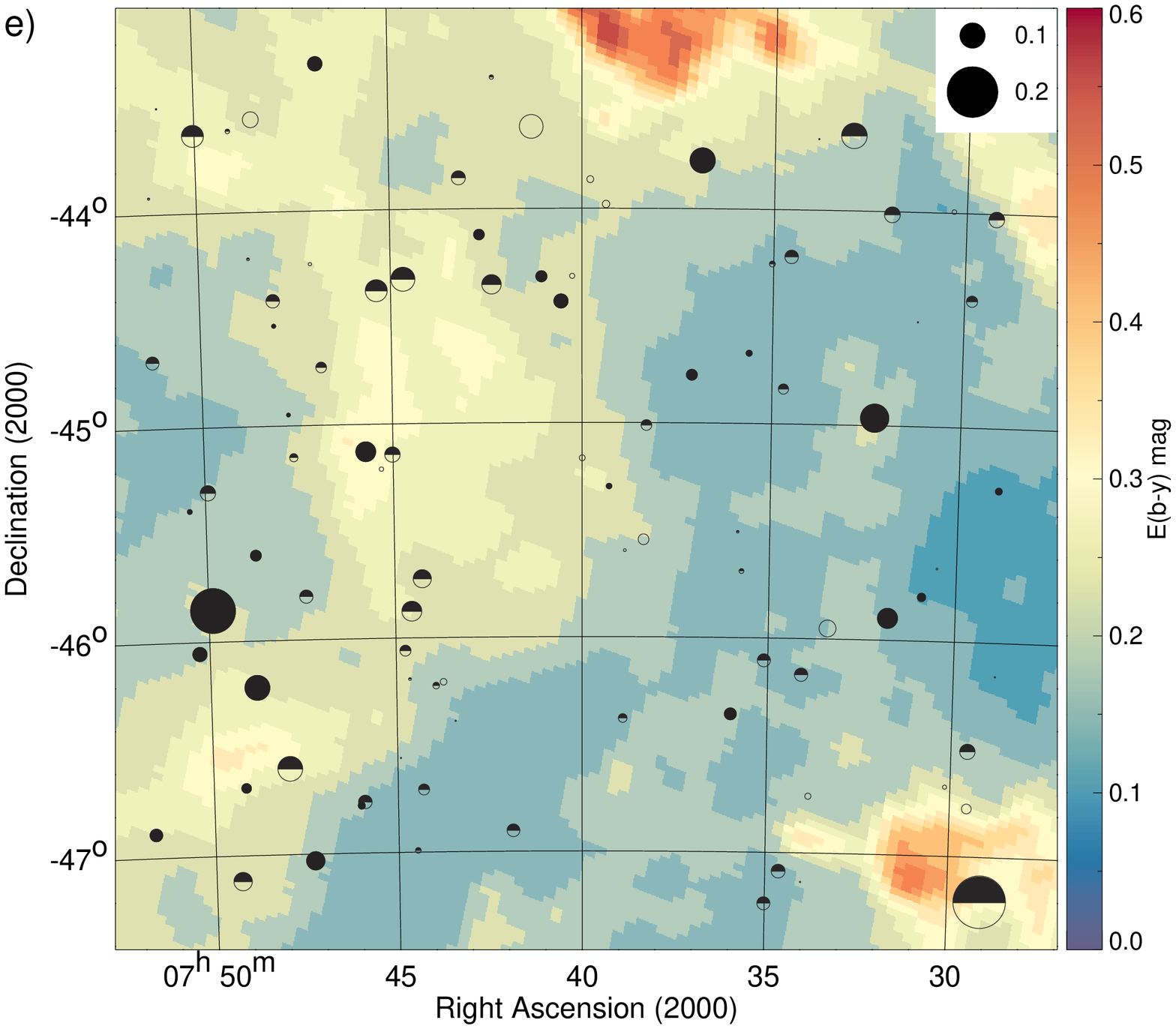}\hskip.20cm\includegraphics[width=.48\hsize]{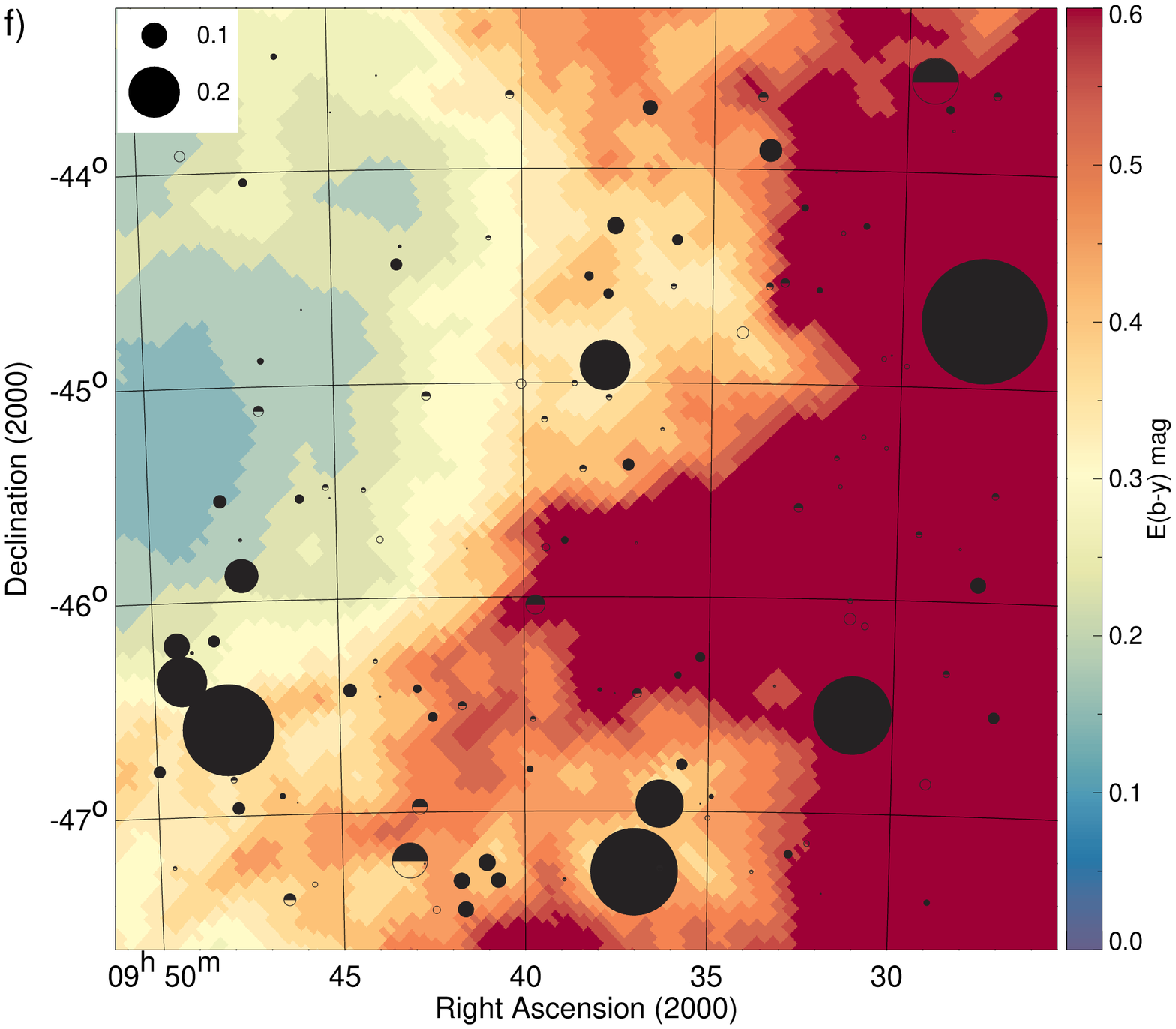}}
%\hbox{\includegraphics[width=.48\hsize]{sfd_147_BW.eps}\hskip.20cm\includegraphics[width=.48\hsize]{sfd_148_BW.eps}}
%\hbox{\includegraphics[width=.48\hsize]{sfd_149_BW.eps}\hskip.20cm\includegraphics[width=.48\hsize]{sfd_170_BW.eps}}
%\hbox{\includegraphics[width=.48\hsize]{sfd_171_BW.eps}\hskip.20cm\includegraphics[width=.48\hsize]{sfd_173_BW.eps}}
\vskip-.8cm
\end{center}
\caption{SFD98 reddening images of the six selected areas: a) SA\,147; b) SA\,148; c) SA\,149; d)
SA\,170; e) SA\,171; and f) SA\,173. The lines-of-sight for which colour excesses have been estimated 
are overlaid upon the images. The sizes of the symbols scale with the estimated value of E$(b-y)$ 
according to the scale shown in the upper corner inset, the negative colour excesses being
assumed to be null. The stellar sample was subdivided into three ranges of distances: $d \le 200$\,pc 
(open circle),  $200 < d \le 400$\,pc (filled upper half circle), and $d > 400$\,pc (filled circle). 
(A colour version of this figure is available in the online journal).}
\label{sfd}
\end{figure*}
%-----------------------------------------------------------------------------

The new distribution of colour excess as a function of the distance for SA\,173 strengthens the
previous result by the inclusion of about 50\% more data points. Although four of the six most 
reddened stars in the observed stellar sample belong to SA\,173, the observed colour excess
towards this line-of-sight is, in general, rather low, averaging $\langle {\rm E}(b-y) \rangle = 
0\fm046$.  SA\,173 is also the only area where stars with colour excess larger than
E$(b-y) = 0\fm3$ have been detected at distances smaller than 1\,kpc. One of these stars is
SA173.0022, which at an estimated distance of $710\pm80$\,pc and E$(b-y) = 0\fm491\pm0\fm007$, 
is the most reddened star in the analysed sample. Two other stars with a rather large
reddening and an estimated distance smaller than 1\,kpc are SA173.0164, at $480\pm90$\,pc and 
E$(b-y) = 0\fm307\pm0\fm006$, and SA173.0970, at $860\pm300$\,pc and 
E$(b-y) = 0\fm359\pm0\fm002$. These rather large reddenings may be related to the interstellar 
material belonging to the outskirts of the VMR. 

As a common characteristic of the observed selected areas, one notes that the lower envelope of 
the derived colour excesses increases with distance at distances greater than 200\,pc from the Sun. 
Nevertheless, some lines-of-sight have quite low colour excesses (E$(b-y) \sim 0\fm05$) even at 
distances larger than $\sim$800\,pc, characterising the low volume density of this region of the sky.

\subsection{A comparison with SFD98's interstellar reddening}\label{sfd98}

An instructive comparison can be made between the estimated {\it uvby--}$\beta$ colour excesses 
for the observed selected areas and the interstellar reddening map obtained by SFD98. This
comparison is depicted in the panels shown in Fig.\,\ref{sfd}, where the estimated colour excess
for the observed stars are overlaid upon the reddening maps. The size of the symbols scales with 
the measured colour excess according to the scale shown in the upper corner of the panels. 
To provide an idea of the distribution of the obtained colour excess as a function of the distance, the 
stellar sample was subdivided into three distance groups, which are represented by different 
symbols: open circles (the nearest group, $d \le 200$\,pc); filled upper half circles (the 
intermediate group, $200 <  d \le 400$\,pc); and filled circles (the farthest group, $d > 400$\,pc).
Moreover, to make the comparison between both results more quantitative, the SFD98 
colour excess, E$(B-V)$, was converted into E$(b-y)$. This conversion was done assuming a
``standard'' interstellar medium, where E$(b-y)$\,$\approx 0.721$\,E$(B-V)$.  

The SFD98 reddening map provides an integrated colour-excess along the line-of-sight, while the 
measured stellar colour-excess provides the reddening out to the star's distance. In addition, many 
authors \citep[e.g.,][]{AG99, DAC03, CLB05} agree that SFD98 may overestimate the reddening 
by a factor of 1.3--1.5 when E$(B-V)$ exceeds values of about 0\fm2--0\fm3. However, a visual 
inspection of Fig.\,\ref{sfd} suggests that there is, in general, a good correlation between 
both estimates of the interstellar absorption towards the analysed areas. 

In particular, we note the good agreement obtained for SA\,170 (Fig.\,\ref{sfd} -- panel ``d''), which is
the observed area with the lowest level of interstellar reddening. When the stellar lines-of-sight are 
used to retrieve the predicted SFD98 colour excess, E$(B-V)$, one obtains values ranging from 
about 0\fm04 to 0\fm12, corresponding to approximately $0\fm03 \le {\rm E}(b-y) \le 0\fm09$. These
values are a few hundredths of magnitude larger than the ones provided by the Str\"omgren
photometry, although this may again be due to the predicted SFD98's reddening providing the total 
column density, while the stellar sample probes the volume out to a distance of about 800\,pc.

Despite the good agreement observed for the surveyed areas, a special word of caution seems 
appropriate regarding SA\,148 and SA\,173 (Fig.\,\ref{sfd} -- panels ``b'' and ``f'', respectively). 
Both areas are located at low Galactic latitudes where most contaminating sources have not been 
removed from SFD98's maps, meaning that their predicted interstellar reddening should not be fully
trusted. Nevertheless, even for these areas the comparison between the estimated {\it uvby--}$\beta$ 
colour excesses and SFD98's reddening seems reasonable. These are the areas where one finds 
the stars with the largest reddening in the whole sample. 

Selected area 171 (Fig.\,\ref{sfd} -- panel ``e'') has the most interesting line-of-sight, not only 
because it is along the IVS, but also because it contains some cometary globules
within its volume. There are twenty-two stars in the nearest group ($d \le 200$\,pc) and the 
estimated colour excess is, in general, smaller than E$(b-y) = 0\fm045$, with only three exceptions. 
One exception is the star SA171.0550 ($\alpha_{2000} = 7^{\mathrm h}41^{\mathrm m} 
18^{\mathrm s}$, $\delta_{2000} = -43\degr 37^{\prime} 05^{\prime\prime}$), which has an excess of 
E$(b-y) = 0\fm093\pm0\fm003$ at an estimated distance of $144\pm42$\,pc. Inspection of the area
(Fig.\,\ref{sfd} -- panel ``e'') shows that this line-of-sight coincides with a portion of relatively large
reddening, suggesting that at least part of the absorption detected by SFD98 in this
direction is due to dust material located nearer than 150\,pc from the Sun. 

\subsection{Cometary globules}

\subsubsection{CG\,4/CG\,6}\label{sec_cg4_6}

\begin{figure}
\centering
\includegraphics[width=\hsize]{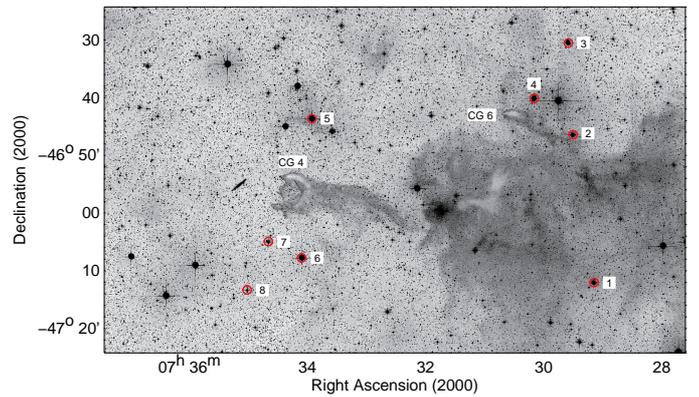}
\caption{Optical (DSS) image of the CG\,4 and CG\,6 region. The estimated data (colour excess 
and distance)
for the identified stars are given in Table\,\ref{cg4_6exc}.}
\label{cg4_6}
\end{figure}

Figure\,\ref{cg4_6} is an optical counterpart detail of the bottom right corner of Fig.\,\ref{sfd} (panel ``e'') 
and displays the region surrounding CG\,4 and CG\,6. The former, which is shown just to the left of the 
centre, has an impressive intricate structure. About half of a degree to the west of CG\,4, there is a 
larger diffuse cloud known as Sa101, and north of this cloud one finds the other cometary globule, CG\,6. 

This region has been identified as a star-forming site. \citet{RP93} found seven H$\alpha$ emission
objects, which are supposedly young stars related to these interstellar structures, and more recently 
\citet{RJH11} extended this number by identifying another 16 candidate young stars.

Colour excesses and distances have been obtained for eight stars in the region, as identified in 
Fig.\,\ref{cg4_6} and shown in Table\,\ref{cg4_6exc}. Among these stars, four are 
supposedly closer than 200\,pc, and have colour excesses of smaller than E$(b-y) \sim 0\fm04$.
The remaining four stars, $d > 200$\,pc, seem to be reddened by rather large colour
excesses, E$(b-y) \ge 0\fm05$. Unfortunately, the object with the largest reddening,
SA171.0025, has a quite large uncertainty in its H$\beta$ measurement ($\sigma_{\rm H\beta}
= \pm0\fm045$), which causes a large inaccuracy in its distance determination. Nevertheless,
the data listed in Table\,\ref{cg4_6exc} suggest that there is a relatively large interstellar absorption
around 200\,pc from the Sun. It is, however, impossible to assure that this increase is related
to the dust associated with CG\,4/Sa101/CG\,6. \citet{RJH11} proposed a yet greater distance 
($\sim$500\,pc) to ensure that the young objects in this region to have age comparable to
those in Taurus. We note, however, that 200\,pc is the distance suggested by \citet{KN00} 
to the cometary globule complex, CG\,30/31/38, which is also a low-mass star-formation site 
located about 12\degr\ to the north of the present complex.

%--------------------------- Figure 6 ----------------------------------------
%-----------------------------------------------------------------------------

\begin{table}
\caption{Selected stars in the region of CG\,4 and CG\,6. The second column gives the
identifying number displayed in Fig.\,\ref{cg4_6}.}
\centering
\begin{tabular}{cccl}
\hline
\noalign{\smallskip}
Star & Num & $E(b-y)$ & \multicolumn{1}{c}{Distance} \\
& &  (mag) &  \multicolumn{1}{c}{(pc)}  \\
\noalign{\smallskip}
\hline
\noalign{\smallskip}
171.~~~25 & 1 & 0.206$\pm0.040$ & $371^{+300}_{-140}$ \smallskip \\
171.~~~39 & 2 & 0.038$\pm0.010$ & $176^{+36}_{-27}$ \smallskip \\
171.~~~38 & 3 & 0.060$\pm0.030$ & $205^{+119}_{-59}$ \smallskip \\
171.~~~59 & 4 & 0.016$\pm0.003$ & $137^{+14}_{-13}$ \smallskip \\
171.~195 & 5 & 0.024$\pm0.002$ & $115^{+15}_{-17}$ \smallskip	 \\
171.~203 & 6 & 0.003$\pm0.004$ & $163^{+22}_{-23}$ \smallskip \\
171.~229 & 7 & 0.054$\pm0.009$ & $293^{+42}_{-35}$ \smallskip \\
171.~246 & 8 & 0.051$\pm0.015$ & $350^{+140}_{-75}$ \smallskip \\
\hline
\end{tabular}
\label{cg4_6exc}
\end{table}

\subsubsection{CG\,5}

\begin{figure}
\centering
\includegraphics[width=\hsize]{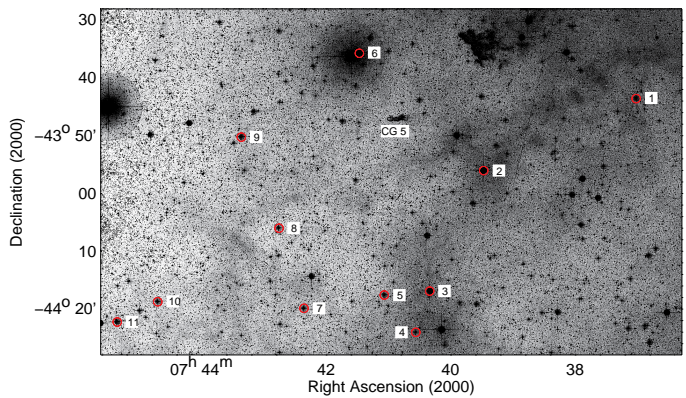}
\caption{Optical (DSS) image of the CG\,5 region. The estimated data (colour excess and distance)
for the identified stars are given in Table\,\ref{cg5exc}.}
\label{cg5}
\end{figure}

CG\,5 is not so impressive as CG\,4, or even CG\,6, and appears to be relatively small in the optical 
image displayed in Fig.\,\ref{cg5}, with dimensions of only about 4\,arcmin. Table\,\ref{cg5exc} gives 
the derived colour excess and distance to eleven stars with lines-of-sight around CG\,5. One of these 
stars is SA171.0550, already discussed in the previous Section, which has a colour excess of almost 
E$(b-y) =0\fm1$ at a distance smaller than 200\,pc. The other two stars closer than that have quite 
small reddenings (E$(b-y) < 0\fm03$), while three stars located at distances ranging from 200\,pc to 
270\,pc have reddenings similar that of SA171.0550 ($0\fm07 \le {\rm E}(b-y) \le 0\fm10$). Once more, 
it seems that there is some interstellar absorption around 200\,pc from the Sun. Nevertheless, it is 
noteworthy mentioning that SFD98 predict quite larger colour excesses towards this line-of-sight than 
those obtained for these stars (see Fig.\,\ref{sfd} -- panel ``e'', upper centre), suggesting that even if 
part of the reddening is produced at distances smaller than 200\,pc, most of it has its origin beyond 
$\approx$1\,kpc. The low density character of this volume is once again testified by the farthest object 
in this region, a B-type star (SA171.0314), which has a quite small reddening (E$(b-y) \approx 0\fm1$) 
for its estimated distance ($d \approx 2$\,kpc).

%--------------------------- Figure 7 ----------------------------------------
%-----------------------------------------------------------------------------

\begin{table}
\caption{Selected stars in the region of CG\,5. The second column gives the
identifying number displayed in Fig.\,\ref{cg5}.
}
\centering
\begin{tabular}{cccl}
\hline
\noalign{\smallskip}
Star & Num & $E(b-y)$ & \multicolumn{1}{c}{Distance} \\
& &  (mag) &  \multicolumn{1}{c}{(pc)}  \\
\noalign{\smallskip}
\hline
\noalign{\smallskip}
171.~314 &  1 & 0.101$\pm0.006$ & 2170$^{+1660}_{-843}$   \smallskip \\
171.~450 &  2 & 0.029$\pm0.005$ &  182$^{+32}_{-25}$  \smallskip \\
171.~497 &  3 & 0.020$\pm0.008$ &  153$^{+49}_{-35}$  \smallskip \\
171.~518 &  4 & 0.058$\pm0.031$ &  504$^{+354}_{-221}$  \smallskip \\
171.~543 &  5 & 0.046$\pm0.016$ &  522$^{+241}_{-156}$  \smallskip \\
171.~550 &  6 & 0.093$\pm0.003$ &  144$^{+42}_{-33}$  \smallskip \\
171.~612 &  7 & 0.076$\pm0.012$ &  232$^{+57}_{-51}$  \smallskip \\
171.~629 &  8 & 0.044$\pm0.003$ &  928$^{+182}_{-183}$  \smallskip \\
171.~643 &  9 & 0.055$\pm0.015$ &  299$^{+196}_{-100}$  \smallskip \\
171.~715 & 10 & 0.094$\pm0.005$ &  267$^{+104}_{-76}$  \smallskip \\
171.~759 & 11 & 0.086$\pm0.010$ &  202$^{+65}_{-51}$   \smallskip \\
\hline
\end{tabular}
\label{cg5exc}
\end{table}

\section{\ion{Na}{i}\,D lines}\label{sodio}

Spectra for six bright early-type stars in the Vela region were gathered as a by-product of 
an observational mission designed to observe the sodium D1 and D2 lines  at 5896 \AA\ and 
5890 \AA\ for stars towards the Southern Coalsack. The observations were performed at 
ESO (La Silla, Chile) in 1989 April by means of the Coud\'e spectrograph (CES) fed by the
1.4\, m telescope. The spectra were extracted and calibrated using standard IRAF routines, 
as described in a previous paper \citep{FR00}. Gaussian fits to the calibration spectral lines 
yielded an actual instrumental resolution (full width at half-maximum, FWHM) of 0.095 \AA\ 
(or 4.8\,km\,s$^{-1}$), corresponding to a resolving power $R \approx 60\,000$. The spectral 
region surrounding the \ion{Na}{i} D lines coincides with a strong concentration of telluric lines, 
notably of atmospheric water vapour. To remove these atmospheric lines from the stellar 
spectra, a synthetic telluric spectrum was constructed using the data listed by \citet{LAML91}, 
that was then convolved with the instrumental resolution of the CES, and scaled to provide 
the corresponding intensity after correction for the air mass of the observations. Applying the
synthetic spectrum proved to produce more accurate results than using the obtained spectrum 
of a lightly reddened early-type star ($\lambda$ Sco) on the same nights that the programme 
stars were observed.

%--------------------------- Figure 8 ----------------------------------------
\begin{figure*}
\centering
\includegraphics[width=.32\hsize]{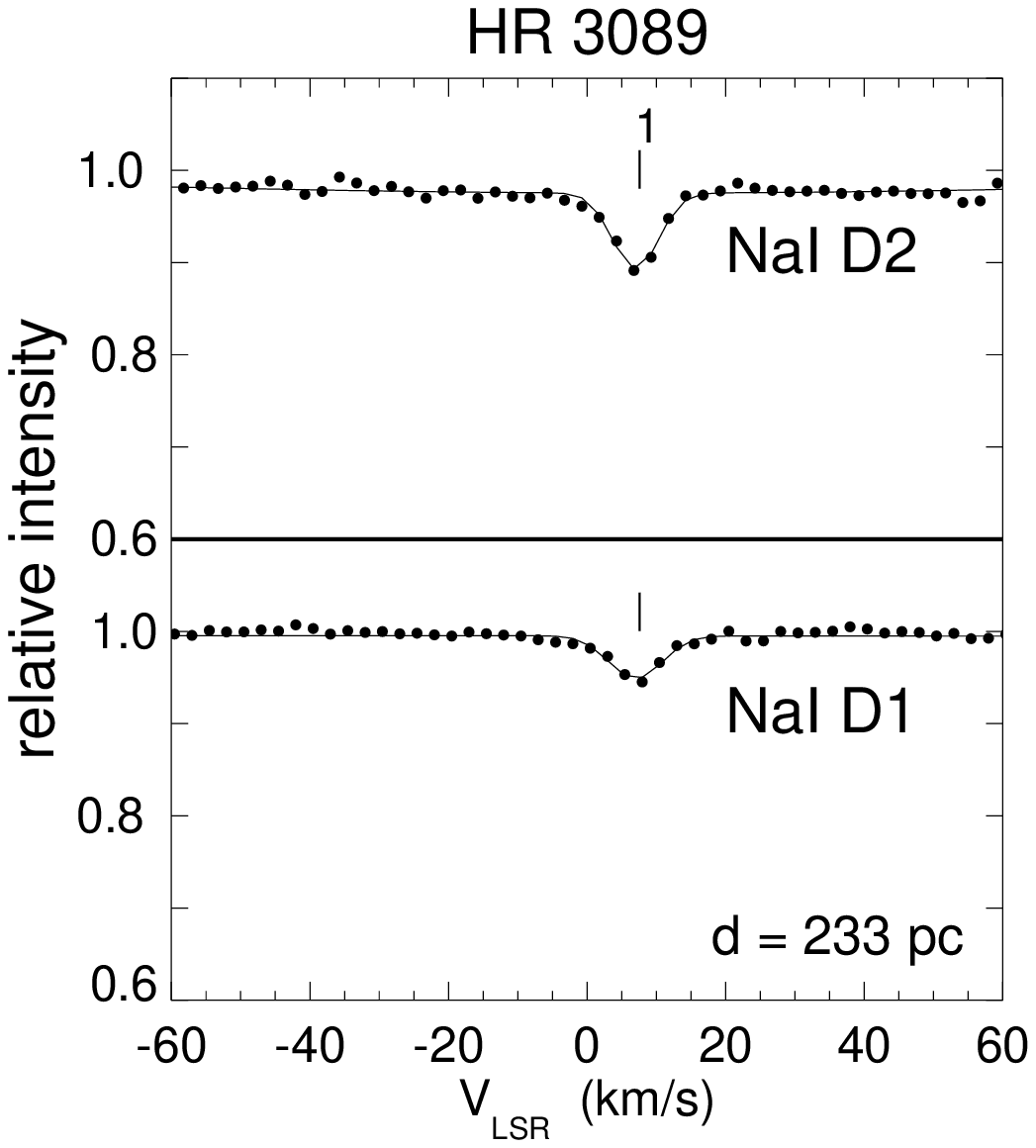}
\includegraphics[width=.32\hsize]{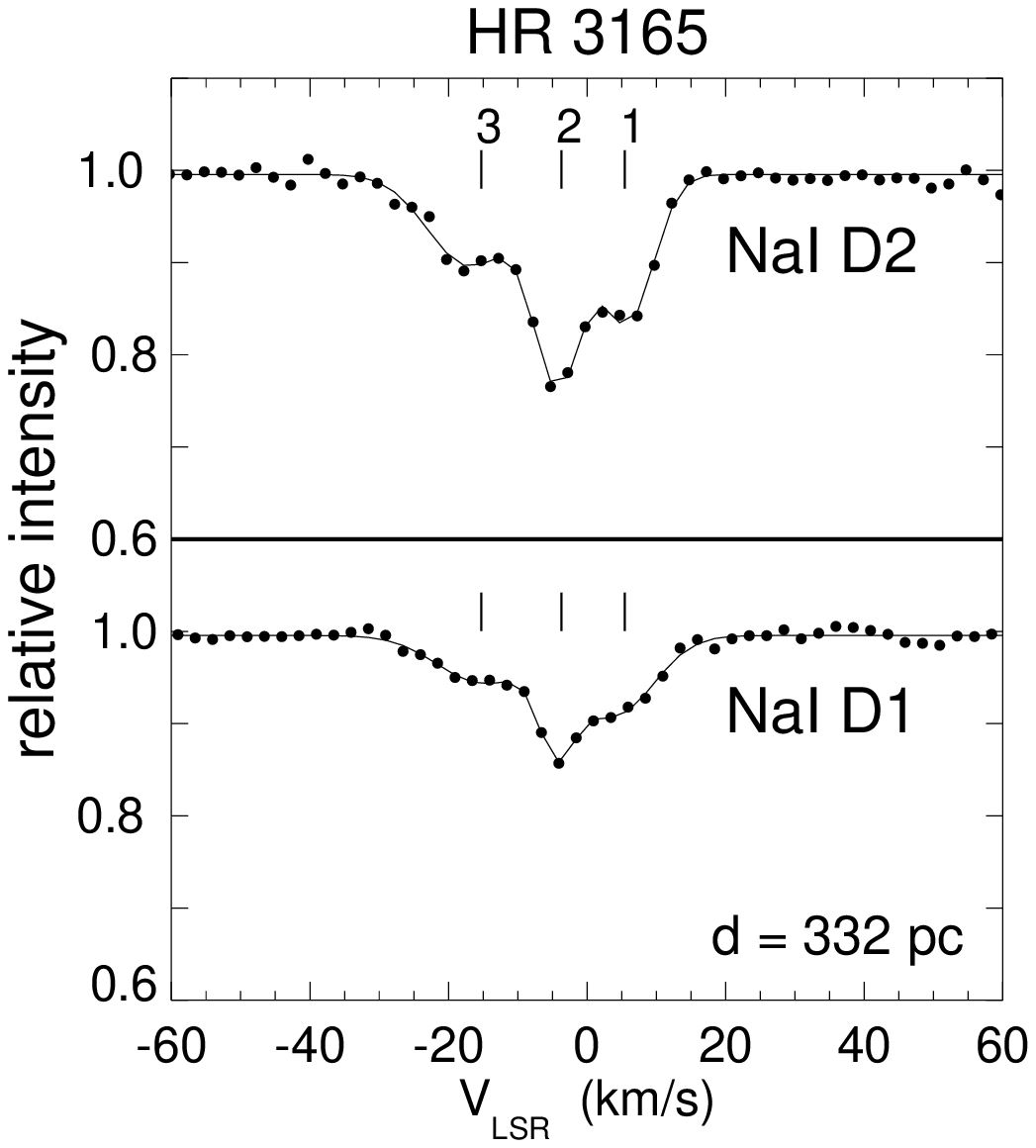}
\includegraphics[width=.32\hsize]{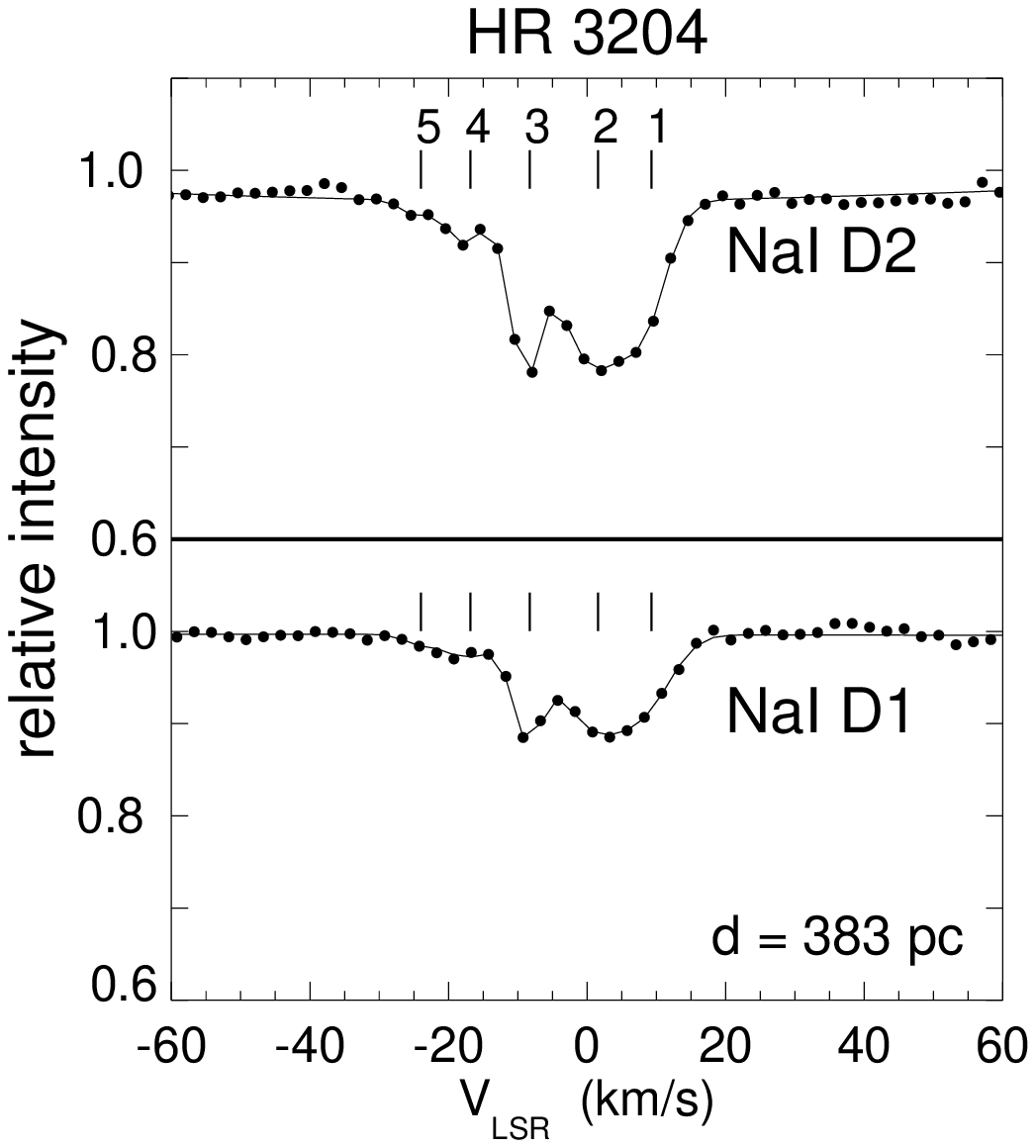}

\includegraphics[width=.32\hsize]{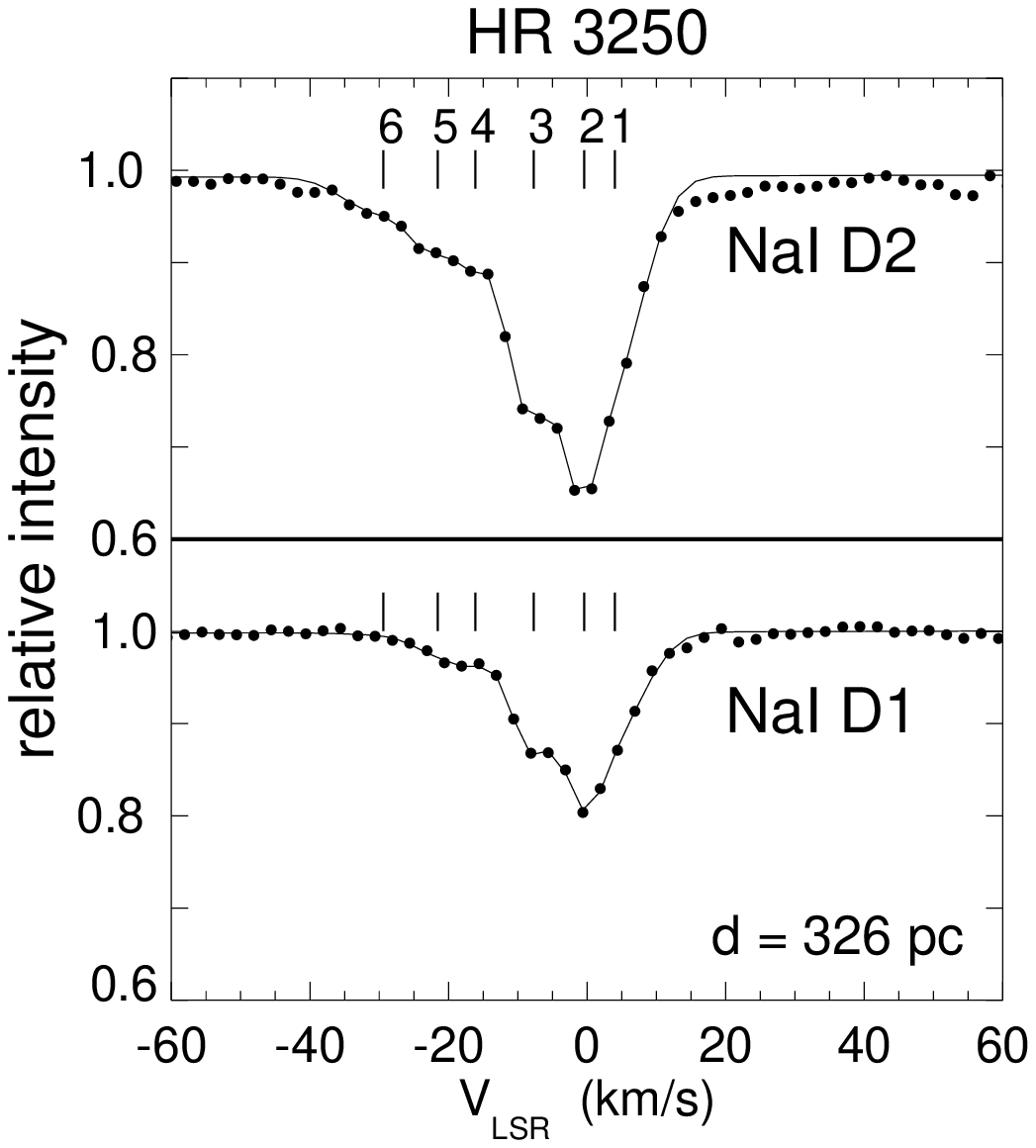}
\includegraphics[width=.32\hsize]{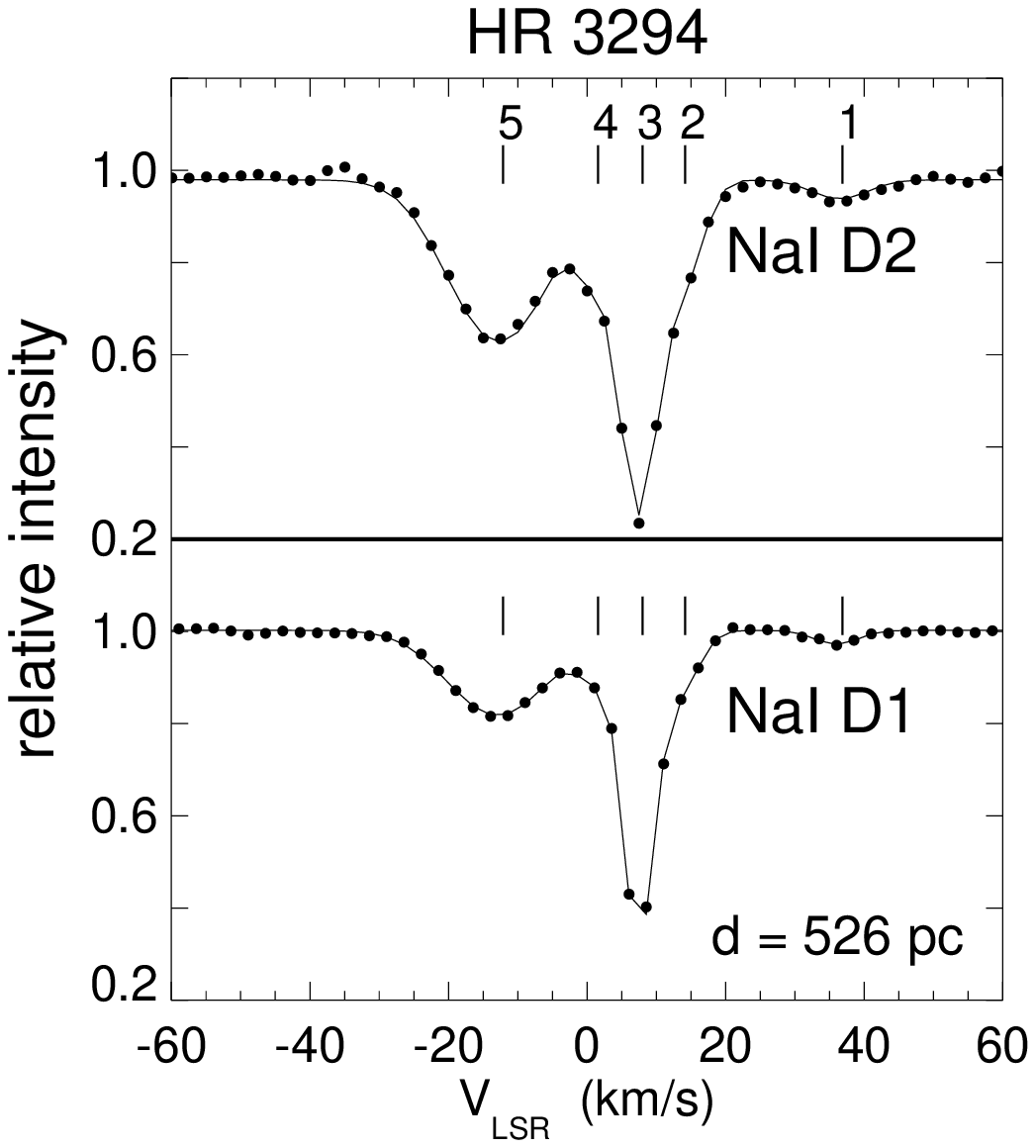}
\includegraphics[width=.32\hsize]{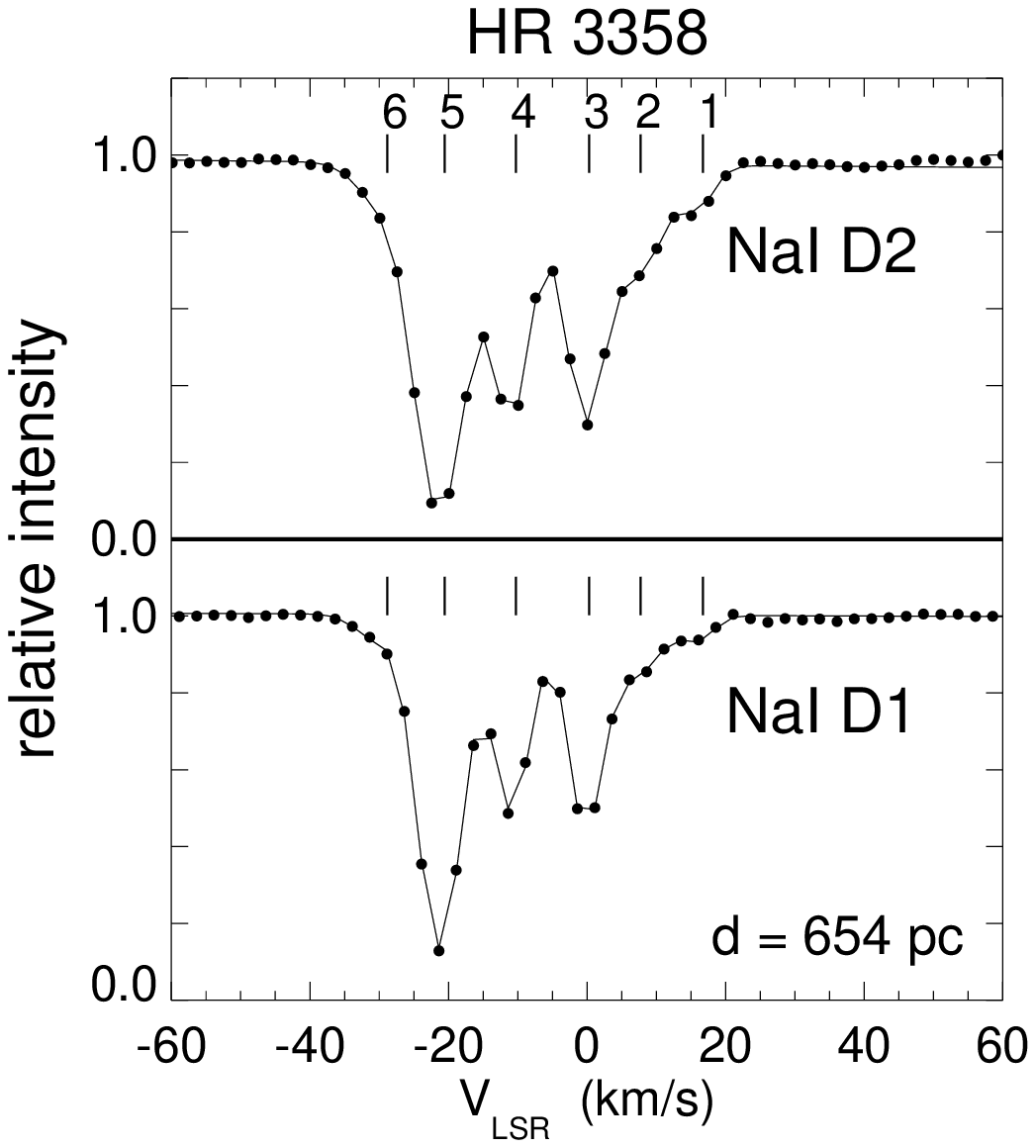}
\caption{Observed normalised interstellar \ion{Na}{i}\,D absorption-line profiles (dots). The 
solid line shows the adopted fitted profile obtained with the parameters given in 
Table\,\ref{nai_par}. The locations of the identified components are indicated by the tick 
marks. Telluric lines have been removed from the normalised profiles by applying a
synthetic telluric spectrum (see text). The new {\it Hipparcos} stellar distance \citep{leeuwen07} is 
given in the right-hand bottom of each panel.}
\label{nai}
\end{figure*}
%-----------------------------------------------------------------------------

Theoretical absorption-line profiles were fitted to the observed spectra using the method described 
by \citet{WHK94}. In this analysis, each cloud component was characterised by a radial velocity, $v$, 
column density of absorbers, $N$, and velocity dispersion parameter, $b$. The calculated profiles 
were convolved with the instrumental resolution to enable comparison with the observations. 
The parameters $v$, $b$ and $N$ were adjusted by trial and error until a satisfactory fit 
was achieved for each one of the \ion{Na}{i}\,D absorption lines. The best obtained fit for 
each star is shown in Fig.\,\ref{nai}, and the derived parameters are given in 
Table\,\ref{nai_par}. The radial velocities are referenced to the Local Standard of Rest (LSR)
frame. To the Sun was assigned a velocity of 16.5\,km\,s$^{-1}$ in the direction 
$l=53\degr$, $b=+25\degr$ as defined by \citet{MB81}, which is in close agreement to
the value more recently obtained by \citet{CAB11} based on the kinematical analysis of almost
20\,000 high-probability thin-disk dwarfs within 600\,pc of the Sun.

HR\,3089 is the nearest of the six observed stars. According to the latest reduction of the 
{\it Hipparcos} astrometric data \citep{leeuwen07}, this star is at a distance of $233^{+8}_{-9}$\,pc. 
In contrast to the observations of the other five stars, the \ion{Na}{i}\,D lines have only one rather 
weak component. It is interesting to compare this result with the one obtained using photometric 
measurements. Adopting the relation between $N$(\ion{Na}{i}) and $N$(H) proposed by 
\citet{FVG85} and the one proposed by \citet{knude78b} to convert $N$(H) into $E(b-y)$, the 
estimated \ion{Na}{i} column density was used to infer a colour excess of $E(b-y) \approx 0\fm002$, 
showing, once again, the low-density volume character towards this direction out to at least 200\,pc. 
The Str\"omgren {\it uvby--}$\beta$ photometric measurements obtained for HR\,3089 
\citep{GO76,GO77}, provide a colour excess E$(b-y) = 0\fm008$ and a distance of 
$378\pm55$\,pc. Both estimates of colour excesses are consistent, however, the estimated 
photometric distance is about 60\% larger than the one obtained by astrometric measurements.

The obtained column density (log N(\ion{Na}{i}) = 10.91\,cm$^{-2}$) and equivalent width 
(W$_{\lambda}$(D$_{2}$) = 13.87\,m\AA), for this star, are slightly smaller than the value usually 
assumed for the ``dense wall'' of neutral gas that surrounds the Local Cavity \citep[e.g.,][]{WLV10}. 
However, from this measurement alone, it is impossible to determine whether this absorption is 
caused by either a large-scale interstellar structure or a single very diffuse cloudlet.

The multi-component \ion{Na}{i}\,D line spectra observed for the remaining five stars attest to
the complexity of the interstellar structure towards the Gum nebula direction. After excluding 
HR\,3165, the other four stars seem to contain at least five or six interstellar components. These
multi-component structure were found by previous investigations of the region 
\citep[e.g.,][]{CSD99,CS00}. 

\begin{table}
\caption{Obtained \ion{Na}{i} absorption-line parameters for the six stars observed 
spectroscopically towards the Vela region.}
\centering
\begin{tabular}{lcrrrrr}
\hline
\noalign{\smallskip}
\multicolumn{1}{c}{HR} & \#& \multicolumn{1}{c}{v$_{lsr}$} & \multicolumn{1}{c}{log\,N} & \multicolumn{1}{c}{b} & \multicolumn{1}{c}{W$_{\rm D1}$} & \multicolumn{1}{c}{W$_{\rm D2}$} \\ 
& & \multicolumn{1}{c}{(\kms )} & \multicolumn{1}{c}{(cm$^{-2}$)} &  \multicolumn{1}{c}{(\kms )} & \multicolumn{1}{c}{(m\AA )} & \multicolumn{1}{c}{(m\AA )} \\ 
\noalign{\smallskip}
\hline
\noalign{\smallskip}
3089 & 1 &     6.9 & 10.91 & 4.2 &  8.89 & 13.87 \\ \\
3165 & 1 &     4.7 & 11.31 & 6.0 & 24.95 & 29.08  \\
         & 2 & $-$4.4 & 11.24 & 3.3 & 12.23 & 39.68 \\
         & 3 & $-$16.1 & 11.21 & 8.3 & 16.89 & 29.44 \\ \\
3204 & 1 &     8.5 & 11.01 & 4.0 & 10.90 & 17.76 \\
         & 2 &     0.8 & 11.38 & 6.1 & 24.48 & 42.10 \\
         & 3 & $-$9.0 & 11.07 & 1.4 & 10.92 & 19.85 \\
         & 4 & $-$17.7 & 10.58 & 3.4 &  4.41 &  6.21 \\
         & 5 & $-$24.8 &  9.93 & 1.0 &  0.84 &  1.67 \\ \\
3250 & 1 &     3.3 & 11.43 & 5.8 & 24.52 & 51.66 \\
         & 2 &  $-$1.2 & 11.24 & 2.6 & 17.66 & 28.63 \\
         & 3 &  $-$8.5 & 11.33 & 3.4 & 18.18 & 42.55 \\
         & 4 &  $-$16.8 & 10.66 & 2.5 & 4.21 & 9.42 \\
         & 5 &  $-$22.3 & 10.64 & 3.5 & 3.37 & 10.49 \\
         & 6 &  $-$30.2 & 10.42 & 6.3 & 0.56 & 9.29 \\ \\
3294 & 1 &  36.1 & 10.65 & 4.3 & 4.63 & 8.30 \\
         & 2 &  13.4 & 11.26 & 2.9 & 19.01 & 29.07 \\
         & 3 &    7.3 & 12.08 & 2.2 & 74.71 & 102.85 \\
         & 4 &    0.8 & 11.03 & 2.0 & 10.88 &   16.90 \\
         & 5\tablefootmark{a} &  $-$12.9 & 11.86 & 9.5 & 65.12 & 128.89 \\ \\
3358 & 1 &  15.9 & 10.92 & 0.4 & 6.86 & 11.97 \\
         & 2 &    7.7 & 11.48 & 4.0 & 30.17 & 48.37 \\
         & 3 & $-$0.2 & 12.10 & 1.8 & 65.85 & 97.37 \\
         & 4 & $-$11.0 & 12.01 & 2.2 & 68.84 & 99.38 \\
         & 5 & $-$21.1 & 12.79 & 2.2 & 134.47 & 161.53 \\
         & 6 & $-$28.4 & 11.05 & 3.5 & 9.66 & 23.68 \\         
         \hline
\end{tabular}

\tablefoottext{a}{Broad component. It may be either due to an unresolved
multi-component interstellar \ion{Na}{i}\,D line or a stellar photospheric line
from a late-type companion.}
\label{nai_par}
\end{table}

\begin{figure}
\centering
\includegraphics[width=\hsize]{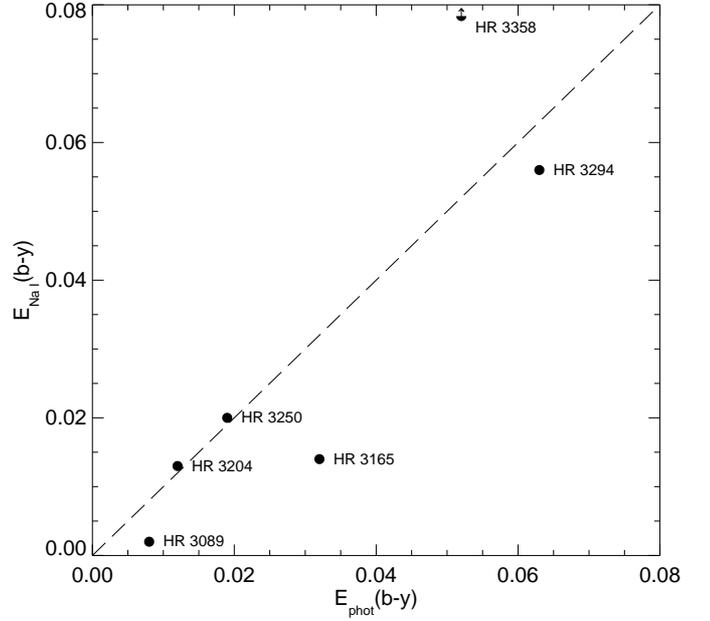}
\caption{Comparison between the estimated colour excess obtained by converting the measured 
\ion{Na}{i} column density into reddening and the value provided by Str\"omgren photometry (see
text for detail). The diagonal dashed-line is shown for reference only.}
\label{na_eby}
\end{figure}

Adopting the aforementioned method to convert a sodium column-density to a colour excess, the 
measured total \ion{Na}{i} column densities for these stars provide values that are in rather good 
agreement with the ones estimated from {\it uvby--}$\beta$ photometry \citep{GO76,GO77}. 
Figure\,\ref{na_eby} compares both results. We note that the measured H$\beta$ value for 
HR\,3165 excludes this object from the range for which the Str\"omgren calibration is valid; for this 
reason, the colour excess was estimated from its $UBV$ photometric data \citep{Co72} and, in the 
case of HR\,3294 the value estimated for the photometric colour excess is only obtained when the 
quite broad component (\#5 in Table\,\ref{nai_par}) is included to compute the total column density. 
For HR\,3358, the estimated total \ion{Na}{i} column density yields a colour excess that is about 
four times larger than the one obtained from photometric measurements. Since the line-of-sight to 
HR\,3358 traverses the Vela SNR, a possible reason for this apparent dust depletion could be 
dust destruction by the action of the supernova's shock wave.

%--------------------------- Figure 9 ----------------------------------------
%------------------------------------------------------------------------------

Some of the fitted components are quite weak, and as for the component obtained for HR\,3089, 
may be associated with the Local Cavity's neutral wall, if it exists towards this line-of-sight. 
For instance, the component \#1 in HR\,3204 has characteristics that are very similar to the one 
observed for the former star. This could also be the case for component \#1 in HR\,3165. 

On the basis of the estimated distances to the six observed stars, it is likely that the material constituting
the front interface of the Gum nebula's expansion shell is acting on the stellar light, except for the case
of HR\,3089, which is supposed to be a foreground object. HR\,3358 may also be affected by the back 
interface. In the non-uniformly expanding model proposed by \citet{WGO01} for the Gum nebula, 
expansion velocities along the line-of-sight within the interval from $-13$ to $+9$\kms are 
expected for the material composing the nebula's shell. The component along the line-of-sight to the 
differential Galactic rotation towards this direction and at a distance ranging from 300\,pc to 500\,pc 
would contributes within the range 1--2\kms. None of the diffuse (log N(\ion{Na}{i}) $ < 11$\,cm$^{-2}$) 
components have velocities within the interval expected for the expansion shell, suggesting that they 
are probably due to single diffuse cloudlets. Some of the ``more dense'' components 
(log N(\ion{Na}{i}) $ > 11$\,cm$^{-2}$) may be associated with the expansion shell, although many of 
them are difficult to explain in a simple scenario. 

Moreover, it is interesting to compare the results obtained for these six stars with the ones 
from previous investigations. \citet{CSD99} and \citet{CS00} analysed the \ion{Ca}{ii} and \ion{Na}{i} 
absorption line profiles of 68 stars in the direction of the Vela SNR. These authors report, for some 
of the observed objects, the detection of high-velocity absorption components and, in some cases, 
line profile variabilities, which were both attributable to the shocked gas associated with the supernova 
remnant. Among the six stars discussed here, either the \ion{Ca}{ii}, \ion{Na}{i}, or both lines of four 
were studied by \citet{CS00}. Although the number of components used to fit the observed spectra are not
exactly the same, in general, the estimated component's velocities agree to within about 1\kms. 
\citet{CS00} report the detection of  weak high-velocity absorption components in the spectra of
three of these four stars. According to those authors, the \ion{Ca}{ii} spectrum gathered for HR\,3250 
(HD\,69302) has a component at $v_{lsr} = +40$\kms, and the one collected for HR\,3358 
(HD\,72108) a component at $v_{lsr} = -111$\kms. None of these components were detected in
the \ion{Na}{i} D spectra introduced here, either because the sodium column densities for these 
components were much smaller than the one for the ionised calcium, or due to the variable character 
observed for some of the line profiles produced by shocked gas. The exception is HR\,3294 
(HD\,70930), for which the reported component at $v_{lsr} = +37$\kms\ observed in their \ion{Ca}{ii}
spectrum corresponds to component \#1 of the \ion{Na}{i} spectrum introduced here. 

\section{Discussion}

\begin{figure}
\centering
\includegraphics[width=\hsize]{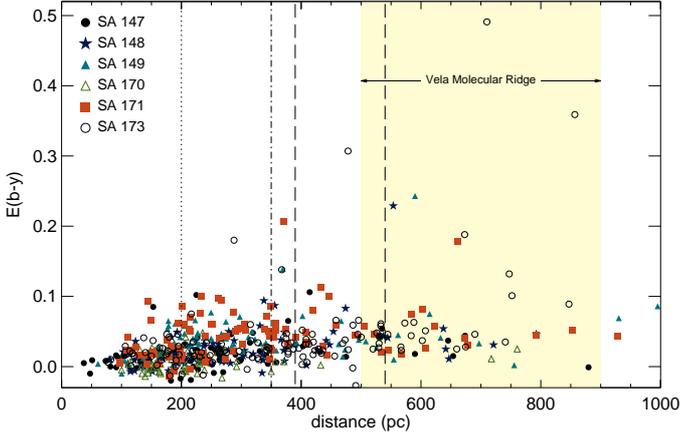}
\caption{Colour excess $versus$ distance diagram obtained when all observed selected areas
are combined. The symbols identifies the selected area from where the data come from, according
to the legend given in the left-hand upper part of the diagram. The vertical lines mark the 
distances proposed in previous investigations for: ($a$) the front face of the Gum nebula 
\citep[dotted line,][]{FR90, KN00}; ($b$) the combined front face of the Gum nebula and IVS 
\citep[dot-dashed line,][]{WGO01}; and ($c$) the front face of the IVS and Gum nebula,
respectively, the left and right lines \citep[dashed lines,][]{SB93}. The coloured band indicates
the range of distances proposed for the Vela Molecular Ridge. (A colour version of this figure is 
available in the online journal).}
\label{gum_total}
\end{figure}

Figure\,\ref{gum_total} gives the obtained colour excess {\it versus} distance diagram when the six
observed selected areas are combined. To make the analysis easier, different symbols were 
used for each selected area. Some interface distances of interest are marked in this diagram. 
Each of them are related to a proposed scenario, as in the case of the dotted line, which gives the 
distance to the front face of the Gum nebula as suggested by \citet{FR90} and \citet{KN00}; the 
dot-dashed line, which indicates the distance to the combined front face of the Gum nebula 
and the IVS proposed by \citet{WGO01}; and the dashed lines that indicate, respectively, the 
distances to the front faces of the IVS and Gum nebula as suggested by \citet{SB93}. The 
coloured band indicates the range of distances proposed for the Vela Molecular Ridge.

The combined diagram shows that very few stars have colour excesses larger than E$(b-y) = 0\fm1$, 
none of which are closer than 200\,pc. The derived mean colour excess for stars out to that distance is 
$\langle {\rm E}(b-y) \rangle \approx 0\fm014$, which corresponds to a hydrogen column-density of 
$N({\rm H}) \approx 10^{20}$\,cm$^{-2}$, that is basically the column density usually associated with 
the Local Cavity. An inspection of Fig.\,\ref{gum_total} suggests that there is a slight rise in the colour 
excess at a distance of $\sim$100\,pc, followed by a steep increase at $\sim$150\,pc, caused by one 
star belonging to SA\,147 and two to SA\,171, the former star being the one already mentioned in 
Sect.\,\ref{colour_dis} for which the parallactic distance is about 250\,pc. Stars belonging to SA\,149
indicate that there is absorbing material at $\sim$180\,pc, but the level of the reddening is smaller than 
that observed for SA\,171.

%--------------------------- Figure 10 ---------------------------------------
%-----------------------------------------------------------------------------

Apart from one star (SA173.0079), the colour excesses do not exceed E$(b-y) \approx 0\fm1$ up
to $\sim$350\,pc. Beyond that distance, one finds three stars with reddenings in the range from
E$(b-y) = 0\fm14$ to $0\fm21$ at basically the same distance, $d \approx 370$\,pc. Nevertheless, 
one of these stars is SA171.0025 (star \#1 in Table\,\ref{cg4_6exc}), which as already mentioned 
in Sect.\,\ref{sec_cg4_6}, has a large uncertainty in its distance determination. The distances to 
the two other stars (SA149.0272 and SA173.0716) are estimated with an accuracy of about 25\%. 
Since each one of these stars is located in a different selected area, it suggests that the 
experienced reddening may be due to an extended layer of absorbing material, which might be 
related to the front interface of the Gum nebula and the IVS, at the distance proposed by 
\citet{WGO01}.

\begin{figure}
\centering
\includegraphics[width=\hsize]{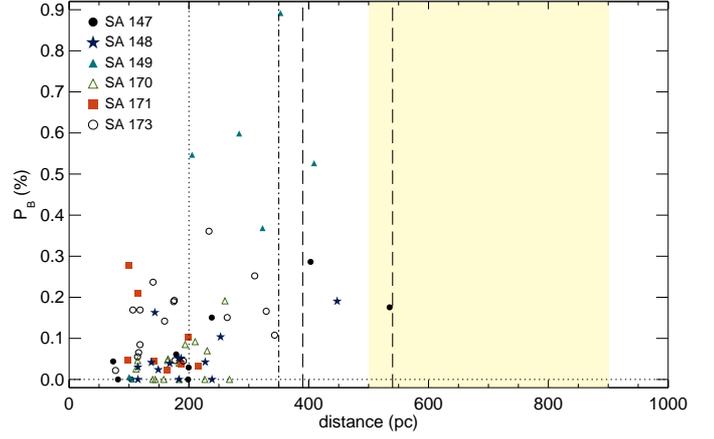}
\caption{Same as Fig.\,\ref{gum_total} for the unbiased $B$-band linear polarisation \citep{FR98a}
$versus$ distance diagram. (A colour version of this figure is available in the online journal).}
\label{polar}
\end{figure}

In an earlier investigation, \citet{FR98a} obtained $UBV$ linear polarisation data for stars distributed 
in 35 selected areas of Kapteyn. There are 71 stars measured polarimetrically in their sample 
belonging to the selected areas analysed here. As expected, the observed degrees of polarisation
for these stars are usually low, although the mean signal-to-noise ratio of the data obtained in
the $B$-band is rather good, $\langle p/\sigma_p\rangle \simeq 4.5$. Figure\,\ref{polar} displays the 
unbiased value of the linear polarisation (i.e., $p_o = 0$ if $p < \sigma_p$ or 
$p_o = (p^2 - \sigma^2_p)^{1/2}$ if $p \ge \sigma_p$), obtained for the $B$-band, $versus$ the distance
to these stars. Most of the stars out to $\sim$200\,pc have virtually null or low ($< 0.1\%$) linear 
polarisation. The exceptions are nine stars belonging to SA\,148 (one star), SA\,171 (two stars), and 
SA\,173 (six stars), among which one of them is SA171.0195 (star \#5 in Table\,\ref{cg4_6exc}). 
Beyond $\sim$250\,pc, the obtained minimum degree of polarisation rises sharply to 
$p_o \approx 0.1\%$, which is a clear indication that there is an extended diffuse interstellar structure 
at a distance of about 200--250\,pc from the Sun towards this region of the sky. We note the quite 
high degree of polarisation presented by the stars that belong to SA\,149 and are located farther away
than $\sim$200\,pc from the Sun. This polarisation is probably due to the interstellar material 
responsible for causing the steep rise observed in the colour excesses of stars beyond 
about 150\,pc (see left middle panel in Fig.\,\ref{excess}). 

Unfortunately, none of the three stars mentioned earlier, that have rather large colour excesses around 
150\,pc (see Fig.\,\ref{gum_total}) have been observed polarimetrically. From the diagram of the colour
excess, the selected area least affected by the interstellar medium is SA\,170, which contains only one 
star with a degree of polarisation higher than 0.1\% and many stars appear to be essentially unpolarised. 

%--------------------------- Figure 11 -------------------------------------
%-----------------------------------------------------------------------------

\section{Conclusions}

We have analysed colour excess and distance obtained for 520 stars distributed in six small areas 
with lines-of-sight towards the Puppis-Vela direction. The main conclusions of this analysis are: 

   \begin{enumerate}
      \item The obtained colour excess {\it versus} distance diagram for stars belonging to these six 
      selected areas confirms the low density character of the probed volume, out to a distance of 
      $\sim$1\,kpc. The estimated colour excess for the analysed stellar sample rarely exceeds 
      E$(b-y) = 0\fm1$.  
      \item Two of the observed selected areas, SA\,147 and SA\,170, have lines-of-sight almost 
      tangent to the proposed rim of the Gum nebula. Most of the surveyed volume within these areas
      is supposed to be outside the nebula. Both volumes have very low levels of reddening, in 
      particular SA\,170, for which the mean colour excess out to a distance of 800\,pc from the Sun is only
      $\langle {\rm E}(b-y) \rangle = 0\fm006$. For SA\,147, the same low reddening was obtained for 
      stars closer than 200\,pc. Beyond that, the reddening appears to increase, indicating that a
      medium with slightly higher density has been reached.    
      \item High resolution \ion{Na}{i}\,D-line spectra obtained for six stars with lines-of-sight towards
      the IVS consists of at least five or six interstellar components out to a distance of
      about 650\,pc from the Sun. The exception is the nearest star, HR\,3089, at a parallactic 
      distance of $\sim$230\,pc, which contains only a single very diffuse component.
      \item Colour excess, polarisation, and absorption \ion{Na}{i}\,D lines, all delineate a diffuse 
      interstellar structure at a distance of about 200--250\,pc from the Sun. The effects of this structure 
      are more clearly evident in the diagram of colour excess $versus$ distance obtained for
      SA\,149 and by the quite high degree of polarisation observed for stars beyond 200\,pc
      in this area.
      \item Although the data clearly show that there is absorbing interstellar material towards 
      the investigated line-of-sight, that may consist of structures at different distances from 
      the Sun, it is impossible to guarantee which one is associated with the Gum nebula's front 
      interface. Colour excesses obtained for three stars suggest that this interface may be at an
      approximate distance of 350\,pc. 
   \end{enumerate}
   
The forthcoming {\it Gaia} satellite, scheduled to be launched in August 2013, will provide accurate
parallactic distances to the stellar sample analysed here. These distances will improve the colour
excess {\it versus} distance diagram and may help us to define more clearly the number of interstellar
structures and their distances from the Sun.

\begin{acknowledgements}
The author thanks the anonymous referee for a constructive and thorough review. The Danish 
Board for Astronomical Research and the ESO (European Southern Observatory) are thanked 
for allocating the observing periods at the Str\"omgren Automatic Telescope 
(Danish 50\,cm). The \ion{Na}{i}\,D lines profile fittings were performed with the help of a computer 
code kindly provided by Dr Daniel E. Welty, who is gratefully acknowledged. The author is grateful 
to the Brazilian agencies CNPq and FAPEMIG for partially supporting this investigation. This research 
has made use of the new $Hipparcos$ catalogue, the Digitized Sky Survey produced at the Space 
Telescope Science Institute under US Government grant NAG W-2166, and the NASA/IPAC Infrared 
Science Archive operated by the Jet Propulsion Laboratory, California Institute of Technology, 
under contract with the National Aeronautics and Space Administration. This research has made 
extensive use of NASA's Astrophysics Data System (NASA/ADS) and the SIMBAD database, operated 
at CDS, Strasbourg, France.    
\end{acknowledgements}

\end{document}